\documentclass[journal,twoside,web]{ieeecolor}
\usepackage{generic}
\usepackage{cite}
\usepackage{amsmath,amssymb,amsfonts}
\allowdisplaybreaks
\usepackage{graphicx}
\usepackage{textcomp}
\usepackage{mathtools}
\usepackage{dsfont}
\usepackage{flushend}
\usepackage{enumerate}
\usepackage{dsfont}
\usepackage{pifont}
\usepackage{algpseudocode}
\usepackage{array}
\usepackage{booktabs}
\usepackage{textcomp}
\usepackage{stfloats}
\usepackage{url}
\usepackage{verbatim}
\usepackage{graphicx}
\usepackage[hidelinks]{hyperref}
\usepackage{subcaption}
\usepackage[affil-it]{authblk}
\usepackage{fancyhdr}
\usepackage{bm}
\usepackage[linesnumbered,ruled,vlined]{algorithm2e}
\SetKwInput{KwInput}{Input}
\SetKwInput{KwOutput}{Output}
\SetKw{KwAssert}{Assert}
\usepackage{centernot}

\usepackage{amsthm}
\newtheorem{lemma}{Lemma}[]
\newtheorem{theorem}{Theorem}[]

\newtheorem{proposition}{Proposition}[]
\newtheorem{corollary}{Corollary}[]
\theoremstyle{definition}
\newtheorem{definition}{Definition}[]
\newtheorem{oldremark}{Remark}[]
\newtheorem{oldexample}{Example}[]

\newtheorem{assumption}{Assumption}[]

\newenvironment{remark}
{\begin{oldremark}\pushQED{\qed}}
	{\popQED\end{oldremark}}
\newenvironment{example}
{\begin{oldexample}\pushQED{\qed}}
	{\popQED\end{oldexample}}

\usepackage{tikz}
\usetikzlibrary{shapes,arrows}
\usetikzlibrary{arrows,calc,positioning}
\tikzset{
	block/.style = {draw, rectangle,
		minimum height=1.2cm,
		minimum width=1.2cm},
	input/.style = {coordinate,node distance=1cm},
	output/.style = {coordinate,node distance=2cm},
	arrow/.style={draw, -latex,node distance=2cm},
	pinstyle/.style = {pin edge={latex-, black,node distance=1cm}},
	sum/.style = {draw, circle, node distance=1cm},
}
\definecolor{backgreen}{HTML}{E9F3DF}
\definecolor{backblue}{HTML}{DAE5EC}
\definecolor{backpurple}{HTML}{E5E0E8}
\definecolor{backorange}{HTML}{F2E9DF}
\definecolor{lightergray}{gray}{0.9}
\definecolor{lightgray}{gray}{0.85}

\usepackage[framemethod=TikZ]{mdframed}
\mdfdefinestyle{callout}{%
	linecolor=black,
	linewidth=1pt,%
	roundcorner=0pt,
	innertopmargin=4pt,
	innerbottommargin=4pt,
	innerrightmargin=5pt,
	innerleftmargin=5pt,
	leftmargin = 0pt,
	rightmargin = 0pt,
	backgroundcolor=lightergray
}

\def\BibTeX{{\rm B\kern-.05em{\sc i\kern-.025em b}\kern-.08em
    T\kern-.1667em\lower.7ex\hbox{E}\kern-.125emX}}
\usepackage{endnotes}

\newcommand{\dint}[0]{\mathrm{d}}

\DeclareMathOperator{\argmax}{\mathrm{argmax}}
\DeclareMathOperator{\EV}{\mathds{E}}

\DeclareMathOperator{\supp}{\mathrm{supp}}

\DeclareMathOperator{\col}{\mathrm{col}}

\DeclareMathOperator{\sign}{\mathrm{sgn}}

\DeclareMathAlphabet{\doublestruck}{U}{BOONDOX-ds}{m}{n}
\newcommand{\ones}[0]{\mathds{1}}
\newcommand{\zeros}[0]{\doublestruck{0}}
\newcommand{\NE}[0]{\mathrm{NE}}

\newcommand{\N}[0]{\mathbb{N}}

\newcommand{\R}[0]{\mathbb{R}}

\newcommand{\Rnn}[0]{\mathbb{R}_{\geq0}}
\newcommand{\Rnp}[0]{\mathbb{R}_{\leq0}}

\newcommand{\Acal}[0]{\mathcal{A}}
\newcommand{\Bcal}[0]{\mathcal{B}}
\newcommand{\Ccal}[0]{\mathcal{C}}

\newcommand{\Fcal}[0]{\mathcal{F}}

\newcommand{\Pcal}[0]{\mathcal{P}}

\newcommand{\Rcal}[0]{\mathcal{R}}
\newcommand{\Scal}[0]{\mathcal{S}}

\newcommand{\Ucal}[0]{\mathcal{U}}

\newcommand{\Xcal}[0]{\mathcal{X}}

\usepackage{xstring}
\newcommand{\weblink}[1]{%
	\StrSubstitute{#1}{https://}{}[\strippedurl]%
	\href{#1}{{\small \color[RGB]{0,0,0} \url{\strippedurl}}}%
}

\newcommand{\muSU}[0]{\mu}

\newcommand{\Rd}[0]{R_\mathrm{d}}

\newcommand{\Rdc}[0]{R_\mathrm{d}^c}
\newcommand{\Rrc}[0]{R_\mathrm{r}^c}

\newcommand{\MSNE}[0]{\mathrm{MSNE}}
\newcommand{\sE}[0]{\mathrm{E}}
\newcommand{\sAE}[0]{\mathrm{AE}}

\newcommand{\sF}[0]{\mathrm{F}}

\newcommand{\aN}[0]{0}
\newcommand{\aL}[0]{\mathrm{L}}
\newcommand{\aH}[0]{\mathrm{H}}

\renewcommand{\baselinestretch}{0.99}

\begin{document}
	\title{Evolutionary Dynamics in Continuous-time Finite-state Mean Field Games - Part~I: Equilibria}
	\author{Leonardo Pedroso, Andrea Agazzi, W.P.M.H. (Maurice) Heemels, Mauro Salazar	\vspace{-1cm}
	\thanks{This work was supported in part by the Eindhoven Artificial Intelligence Systems Institute (EAISI).}
	\thanks{L.~Pedroso, W.P.M.H.~Heemels and M.~Salazar are with the Control Systems Technology section, Department of Mechanical Engineering, Eindhoven University of Technology, The Netherlands (e-mail: \{l.pedroso,m.heemels,m.r.u.salazar\}@tue.nl).}
	\thanks{A.~Agazzi is with the Institute of Mathematical Statistics and Actuarial Science, Department of Mathematics and Statistics, University of Bern, Switzerland (e-mail: andrea.agazzi@unibe.ch).}
	\thanks{See Part~II for the full authors’ biographies.}}
	
%	Evolutionary Dynamics in Continuous-time Finite-state Mean Field Games - Part~I: Equilibria
%	Evolutionary Analysis of Continuous-time Finite-state Mean Field Games under Average Payoffs - Part~I: Equilibria
	\maketitle
	
	\vspace{-0.4cm}
	
	\begin{abstract} %1200 ch
		We study a dynamic game with a large population of players who choose actions from a finite set in continuous time. Each player has a state in a finite state space that evolves stochastically with their actions. A player’s reward depends not only on their own state and action but also on the distribution of states and actions across the population, capturing effects such as congestion in traffic networks. While prior work in evolutionary game theory has primarily focused on static games without individual player state dynamics, we present the first comprehensive evolutionary analysis of such dynamic games. We propose an evolutionary model together with a mean field approximation of the finite-population game and establish strong approximation guarantees. We show that standard solution concepts for dynamic games lack an evolutionary interpretation, and we propose a new concept -- the Mixed Stationary Nash Equilibrium (MSNE) -- which admits one. We analyze the relationship between MSNE and the rest points of the mean field evolutionary model and study the evolutionary stability of MSNE.
	\end{abstract}

% which can capture effects such as congestion. 

%	We consider a model of a large population of players that choose actions from a finite action-space in continuous-time. Each player is characterized by a state in a finite state-space that evolves stochastically as the player chooses actions.  Upon choosing an action, the reward of a player is coupled with the states and actions of the remainder of the players in the population. In the literature, the evolutionary analysis of games is almost solely focused on static games, where players do not have individual states. In this work, for the first time in the literature, we perform a thorough evolutionary analysis of this class of dynamic games. First, we propose an evolutionary model and a mean-field approximation with strong approximation properties in relation to the finite population game. Second, we show that state-of-the-art solution concepts for this class of dynamic games do not have an evolutionary interpretation. We propose a novel solution concept that does, which we call MSNE. Third, we analyze the relation between the novel MSNE solution concept and the equilibria of the mean-field evolutionary model. Fourth, we analyze the evolutionary stability properties of MSNE. 
	
	\begin{IEEEkeywords}
		Stochastic dynamic games, Evolutionary game theory, Mean field games, Nash equilibria, Population games
		%Enter key words or phrases in alphabetical order, separated by commas. Using the IEEE Thesaurus can help you find the best standardized keywords to fit your article. Use the thesaurus access request form for free access to the IEEE Thesaurus: \underline{https://www.ieee.org/publications/services/thesaurus-acce}\\
		%\underline{ss-page.com.}
	\end{IEEEkeywords}

%Dynamic games
%Stochastic dynamic games
%Evolutionary game theory
%Mean field games
%Markov decision processes
%Population games
%Nash equilibria
%Stationary equilibria

\vspace{-0.2cm}

% !TeX spellcheck = en_US
\section{Introduction}

Many interesting systems across diverse disciplines can be modeled by a large population of interacting players (also called agents). These models are relevant, e.g., in economics, where a large number of firms compete and collude in a market \cite{Lambson1984,WeintraubBenkardEtAl2008}; in biology, where animals compete and cooperate for survival of the species \cite{LeimarMcNamara2023}; in engineering, where robots in a swarm cooperate to achieve tasks beyond the capabilities of a single robot  \cite{PedrosoBatistaEtAl2025ULSS}; and in societal studies to predict mobility patterns \cite{Wardrop1952} and investigate opinion dynamics \cite{UrenaKouEtAl2019}. 

In this paper, we consider a model describing such a large population of interacting players. Each player repeatedly chooses an \emph{action} in \emph{continuous time} whenever an individual Poisson clock rings. Players' clocks are independent so players' actions are \emph{asynchronous}.  Each player is characterized by a \emph{discrete state} in a \emph{finite state space}, which evolves stochastically each time a player chooses an action. The set of actions available to a player depends on their state. The \emph{immediate reward} of a player choosing a certain action depends on their state and the state-action distribution across the population, which couples the players' decisions. This can capture effects such as congestion in a traffic network, or the increased attractiveness and cost-efficiency of a firm as more users choose it in an economic setting. Games of this class are called \emph{continuous-time finite-state stochastic dynamic games of many players} \cite{AdlakhaJohariEtAl2015}. In other fields, these are also called \emph{stochastic dynamic games with mean field coupling} and \emph{interacting controlled Markov chains}. Crucially, these games are said to be \emph{dynamic} because players make multiple decisions over time, with each decision (possibly) triggering a change in an individual state of the player. Thus, a player’s decisions affect not only their immediate reward but also the state to which they transition, which in turn shapes the reward and action space at subsequent decision instants. In contrast, we speak of a \emph{static} game of many players (also known as a population game \cite[Chap.~1]{Sandholm2010}) when players do not possess an individual state that influences their reward or action space.

\begin{example}\label{eg:token}
	One motivating application for the analysis carried out in this paper are token economies where a large number of users compete for access to shared resources \cite{PedrosoHeemelsEtAl2023KarmaParallel,PedrosoAgazziEtAl2024EqtEql}. To achieve a fair system-optimal resource allocation, an incentive scheme based on tokens that cannot be traded or bought for money can be used. Each user is provided with a wallet of tokens, which are earned and spent by using the resources. In these settings: (i)~each user makes decisions in \emph{continuous-time} to use a set of resources that satisfy their needs; (ii)~the users' decisions are \emph{asynchronous}, since their need to use the resources is typically uncoordinated and intermittent; (iii)~each user can be characterized by a \emph{discrete state} that is the amount of tokens that they possess and that evolves in jumps as they make decisions to use resources; (iv)~the reward perceived by a user when using the resource depends on the congestion of that resource, which couples the users' decisions. The model for token economies falls into the class of dynamic games which is analyzed in this paper.
\end{example}

In the context of game theory, an important role is played by \emph{solution concepts}, which are rules or criteria that characterize reasonable outcomes of a given game,\footnote{Throughout this paper, the term \emph{solution concept} should not be confused with the term \emph{solution}, which will be used to describe a function of time that satisfies a given differential equation.} such as the celebrated Nash equilibrium (NE). The goal of this paper is to study \emph{solution concepts} for continuous-time finite-state asynchronous stochastic dynamic games of many players, in an attempt to describe its outcome. In particular, we resort to a mean field approximation of the game, i.e., the limit case where the population is infinite, each player carries infinitesimal weight, and each player's single-stage reward is coupled with the mean field of the population rather than individually with all the other players. Mean field-like approaches, introduced in the context of traffic engineering in~\cite{Wardrop1952}, have good approximation properties w.r.t.\ the finite population game, and crucially allow for a tractable analysis of solution concepts~\cite{HaurieMarcotte1985,Borkar2009}. The literature on the analysis of this and similar classes of games is discussed in detail in Section~\ref{sec:sota} below.

Evolutionary game theory was introduced by Smith and Price in the early 1970s for biological modeling \cite{SmithPrice1973,Smith1982}. Since then it has been used to analyze \emph{static games} beyond the classical concept of NE by softening assumptions of rationality, knowledge of the game, and knowledge of the equilibrium by the players \cite{Sandholm2010,ArcakMartins2021}. Indeed, evolutionary game theory introduces a model for the way individual players update their decisions, called \emph{revision protocols}, which are simple myopic rules that, according to some information structure, model how players switch decisions as the mean field and payoffs evolve. In a static game, the player's decisions are their \emph{actions}. Hence, if players are allowed to unilaterally revise their decisions at a given rate according to a specified revision protocol, this induces a time evolution of their actions known as \emph{revision dynamics}. The model of the game, together with these revision dynamics, defines an \emph{evolutionary model}, for which a natural \emph{solution concept} is the stationarity of the revision dynamics. That is, a reasonable outcome of the game is a mean field action distribution that is a rest point of the revision dynamics, which is a point where the proportion of the population choosing each action remains constant in time. Moreover, the evolutionary stability of such points can be assessed by checking if they are immune to mutations of a small fraction of the population.

% \rev{In a static game, the player's decisions are their \emph{randomization of actions} (which can also deterministically be always the same action). Hence, if players are allowed to unilaterally revise their decisions at a given rate according to a specified revision protocol, this induces a time evolution of their randomization of actions} known as \emph{revision dynamics}. \rev{This is called an \emph{evolutionary model}. Under this model, a natural \emph{solution concept} is the stationarity of the revision dynamics. That is, a reasonable outcome of the game is a \rev{mean field} action distribution that is a rest point of the revision dynamics}, \rev{which is} a point where the proportion of the population choosing each action \rev{remains constant in time}. Moreover, the evolutionary stability of such points can be assessed by checking if they are immune to mutations of a small fraction of the population.

\begin{mdframed}[style=callout]
		A solution concept grounded in an \emph{evolutionary} model offers a more compelling notion of a game's outcome than one that is not, such as the NE, because it offers insight into how the outcome emerges under very limited assumptions about the players' knowledge.  In \emph{static games}, the rest points of the revision dynamics -- the natural evolutionary solution concept -- are NE for most meaningful families of revision protocols \cite{Sandholm2010}. In contrast, in \emph{dynamic games}, state-of-the-art solution concepts cannot be given an evolutionary interpretation. The goal of this paper is to initiate a formal \emph{evolutionary analysis of dynamic games}.
\end{mdframed}

%The motivation behind this paper stems from the lack of an \emph{evolutionary analysis} of solution concepts for the class of continuous-time finite-state asynchronous stochastic dynamic games of many players.

We focus our attention on continuous-time finite-state stochastic dynamic games of many players, which is a subclass of the larger domain of dynamic games. In brief, we propose and thoroughly analyze an evolutionary solution concept for the mean field approximation of this class of dynamic games. The proposal we make is in line with the evolutionary game theory literature for static games. We conclude that the proposed solution concept has an evolutionary interpretation according to the proposed evolutionary model and we study its evolutionary stability.

\subsection{State-of-the-art}\label{sec:sota}

The analysis of solution concepts for large population stochastic dynamic games has been extensively analyzed in the literature. There are many meaningful variations of such games which typically have fundamentally distinct properties. The most common defining features of these models are: (i)~the nature of the state and action sets (e.g., finite, countably infinite, or uncountable); (ii)~the timing of the players' decisions (continuous- or discrete-time); (iii)~the nature of the payoff perceived by the players (e.g., infinite- or finite-horizon and discounted or undiscounted).  The analysis of discrete-time finite-state mean field-like games was initiated by \cite{JovanovicRosenthal1988} in the 1980s, which were termed \emph{anonymous sequential games}. In the 2000s, the term mean field game was coined and their study for continuous-time players' decisions \cite{LasryLions2006,LasryLions2006b,LasryLions2007,HuangMalhameEtAl2006} was initiated. Table~\ref{tab:sota_charact} shows a brief characterization of state-of-the-art approaches to dynamic games of many players (for an in-depth survey see \cite{GomesSaude2014,CarmonaDelarue2018}). Even though discrete-time finite-state and continuous-time finite-state games generate a discrete-time and a continuous-time evolution of the mean field, respectively, the principles employed to define solution concepts are similar.

\begin{table*}[ht]
	\centering
	\small
	\vspace{-0.2cm}
	\caption{\small Characterization of state-of-the-art approaches to dynamic games of many players.}\label{tab:sota_charact}
	\vspace{-0.1cm}
	\begin{tabular}{m{2cm}m{3cm}m{3cm}m{4cm}}
		\toprule 
		& State/Action sets &  Players' decisions & Players' payoff  \\
		\toprule
		\cite{LasryLions2006,LasryLions2006b,LasryLions2007}  & Euclidean spaces  & Continuous-time & Average  \\
		\cite{HuangMalhameEtAl2006}  & Euclidean state space, compact action space  & Continuous-time & Undiscounted finite horizon  \\
		\midrule 
		\cite{JovanovicRosenthal1988,BerginBernhardt1992,BerginBernhardt1995} & Compact metric spaces & Discrete-time & Discounted infinite horizon \\
		\cite{SaldiBasarEtAl2018,SaldiBasarEtAl2019} & Polish spaces &  Discrete-time & Discounted infinite horizon \\
		\cite{AdlakhaJohariEtAl2015}  & Countable state space, Euclidean action space   & Discrete-time & Discounted infinite horizon  \\
		\cite{GomesMohrEtAl2010} & Finite spaces & Discrete-time  &  Undiscounted finite horizon\\
		\cite{ElokdaBolognaniEtAl2024} & Finite spaces   & Discrete-time & Discounted infinite horizon \\
		\cite{WiecekAltman2015} & Finite spaces  & Discrete-time & Average and total \\
		\cite{Wiecek2020} & Compact metric spaces  & Discrete-time & Average \\
		\midrule
		\cite{GomesMohrEtAl2013,BayraktarCohen2018} & Finite spaces   & Continuous-time & Undiscounted finite horizon \\
		\cite{DoncelGastEtAl2019} & Finite spaces  & Continuous-time & Discounted infinite horizon \\
		\midrule 
		Our work &  Finite spaces & Continuous-time &  Average \\
		\bottomrule
	\end{tabular}
	\vspace{-0.2cm}
\end{table*}

State-of-the-art solution concepts rely on the notion of a \emph{policy}, which is a map from the state space to a probability distribution over the action space. In \emph{dynamic} games, the concept of a policy is fundamental because it allows players to choose an action depending on how their current state affects future decisions. In contrast, the concept of a policy is moot in \emph{static} games, since a player's action does not influence their future choices. Thus, a player's decision in a static game is characterized solely by an action (which can be interpreted as a degenerate policy with a single state). Accordingly, evolutionary models for static games model how players revise their actions, whereas in dynamic games they must \emph{describe how players revise their policies}. 

However, all the references with finite state space in Table~\ref{tab:sota_charact} rely on solution concepts whereby all players use the \emph{same policy} and no player can unilaterally switch to another policy to increase their payoff. We refer to this as a \emph{behavioral equilibrium}. Intuitively, one can already tell that such a behavioral notion of equilibrium lacks an evolutionary interpretation because it does not allow players to revise their policies \emph{individually}.

\begin{mdframed}[style=callout]
In Section~\ref{sec:equilibria}, we formally define \emph{behavioral} equilibria and show in Section~\ref{sec:ev_dynamics} that they do not have an evolutionary interpretation. The key reason is that there is no room for heterogeneity in the players' behavior, which is essential for defining individual revision dynamics. To address this, we propose a \emph{mixed} solution concept that allows for such heterogeneity. To the best of our knowledge, such a solution concept has not been studied for the class of dynamic games considered in this paper. 
\end{mdframed}

%It is worth remarking that the reason for the popularity of a behavioral solution concept is the ease to establish existence guarantees and to numerically compute it.

Regarding dynamic games, there are only three works \cite{AltmanHayel2010,FleschParthasarathyEtAl2013,BrunettiHayelEtAl2018} that attempt a formal evolutionary analysis. However, all three works consider a setting where players have many random asynchronous pairwise interactions, where the immediate reward of two interacting players depends only on their states and actions (and not on the mean field), which severely limits the generality of the model. In that case, it can be shown that the expected immediate reward depends linearly on the mean field state-action measure (which only captures the probability of interacting with a player with a given state-action pair), whereas the setting under consideration in this paper does not make any assumptions on the dependence of the immediate reward on the mean field. First, in \cite{AltmanHayel2010}, by neglecting the effect of state dynamics in the player's matching probabilities, the authors define an evolutionary stability condition for NE. However, therein, the solution concept is not rooted in revision dynamics of individual players. Second, \cite{FleschParthasarathyEtAl2013} considers a pairwise interaction model with average payoffs that is very similar to the one in \cite{AltmanHayel2010}, differing only in the fact that the state transition probabilities depend on the actions chosen by both interacting players. In \cite{FleschParthasarathyEtAl2013}, a particular revision protocol (called replicator dynamic) is extended to the modeling framework therein by assuming that the players always choose the best-reply policy to the induced stationary population policy. This approach is not in line with the principle behind replicator dynamics in the evolutionary game theory literature, whereby players can possibly myopically switch to non-optimal policies. In \cite{BrunettiHayelEtAl2018}, the authors follow an approach closer to the one proposed in this paper by modeling, under average payoffs, replicator revision dynamics for the players' policies (i.e., players' revise their state-action maps), in line with the myopic principles of evolutionary game theory literature. However, in \cite{BrunettiHayelEtAl2018}, despite modeling coupled state and policy revision dynamics, it is assumed that when a player's clock rings that player's state and policy are drawn at random from the marginal state and marginal policy distributions. Under this approximation, a model is only needed for the marginal state and policy distributions, but it comes at the expense of the loss of an individual model for the players that is consistent as time evolves. One can only argue that this is a valid approximation in case the revision dynamics are orders of magnitude faster than the state dynamics, an assumption that \cite{BrunettiHayelEtAl2018} implicitly leverages to establish results. In this paper, we do not make this approximation. Indeed, we model the evolution of a state-policy joint distribution rather than separate marginal state and marginal policy distributions as in \cite{BrunettiHayelEtAl2018}.

Despite not being exactly aligned with the setting of this paper, recent work has analyzed more robust notions of equilibria through a stability analysis, focusing on static games where the payoff map is dynamic rather than memoryless \cite{FoxShamma2013,ParkMartinsEtAl2019,ArcakMartins2021,MartinsCertorioEtAl2024}.

%\vspace{-0.3cm}

\subsection{Statement of Contributions}
In conclusion, to the best of our knowledge, an evolutionary analysis of mean field games where players' policies are modeled individually has not been studied in the literature. This work fills that gap for continuous-time finite-state mean field games with average payoffs. Specifically, the main contributions of this two-part paper are as follows:
	\begin{itemize}
		\item In Section~\ref{sec:equilibria}, we show that state-of-the-art solution concepts based on a notion of a stationary behavioral Nash equilibrium \emph{lack an evolutionary interpretation}. Therefore, we introduce a novel equilibrium notion for this class of games, the \emph{mixed stationary Nash equilibrium} (MSNE), which admits one. We study its \emph{existence}, \emph{uniqueness}, and \emph{approximation} w.r.t.\ the analogous $N$-player game as $N\to \infty$.
		
		\item In Section~\ref{sec:ev_model}, we formulate an explicit \emph{mean field evolutionary model of the dynamic game} for the first time in the literature. We show that its trajectories approximate those of the analogous $N$-player game as $N\to \infty$. 
		\item In Section~\ref{sec:rest_point_MSNE}, we study the relationship between MSNE and the rest points of the proposed evolutionary model. We establish an equivalence between them for broad classes of meaningful revision protocols.
		\item In Part~II \cite{PedrosoAgazziEtAl2025MFGAvgII} of this work, we investigate the \emph{evolutionary stability} of MSNE. Specifically, we provide conditions on both the MSNE characteristics and the payoff structure of the game under which local and global evolutionary stability results can be established.
	\end{itemize}

%\subsection{Organization}
%
%Section~\ref{sec:model} introduces the finite-population and the mean field models of the game. In Section~\ref{sec:equilibria}, we propose the MSNE solution concept, compare it to alternative state-of-the-art solution concepts, and analyze its existence, uniqueness, and approximation properties. In Section~\ref{sec:ev_model}, we propose finite-population and mean field evolutionary models and establish approximation properties. Section~\ref{sec:rest_point_MSNE} analyzes the relation between MSNE and rest points of the evolutionary dynamics. Section~\ref{sec:MAC_equilibria} illustrates the equilibrium concepts for a well-know real-life example. In Section~\ref{sec:local_ev_stability}, we establish stability properties of MSNE in different settings. Section~\ref{sec:MAC} illustrates the stability concepts and results of this paper for a well-known real-life example. Finally, Section~\ref{sec:conclusion} draws the main conclusions of this paper.

\vspace{-0.3cm}

\subsection{Notation}	
For $N\in \N$, the set of consecutive positive integer numbers $\{1,2,\ldots, N\}$ is denoted by $[N]$.
The $i$th entry of a vector $x\in \R^n$ is denoted by $x_i$. The Euclidean norm of a vector $x\in \R^n$ is denoted by $||x||$. The $n$ dimensional vector of zeros and ones are denoted by $\zeros_{n}$ and $\ones_n$, respectively. Alternatively, $\zeros$ and $\ones$ denote the vectors of zeros and ones of appropriate dimensions, respectively. The sign of $x\in \R$ is denoted by $\sign(x)$ and takes the values of $-1$, $0$, or $1$ if $x<0$, $x=0$, or $x>0$, respectively. The column-wise concatenation of a finite number of vectors $x^1, x^2, \ldots, x^K$ is denoted by $\col(x^1, x^2, \ldots, x^K)$. The indicator function of $a\in \Xcal$ is denoted by $\delta_a:\Xcal \to \{0,1\}$ and $\delta_a(x) = 0$ if $x\neq a$ and $\delta_a(x) = 1$ if $x = a$. The support of a function $f:\Xcal\to \R$ is denoted by $\supp(f):=\{x\in \Xcal: f(x)\neq 0\}$. 
The interior of a set $\Acal$ is denoted by $\mathrm{int}(A)$. Given sets $\Xcal_1,\Xcal_2,  \ldots,  \Xcal_K$,  the Cartesian product $\Xcal_1 \times \Xcal_2 \times \cdots \times \Xcal_K$ is denoted by $\bigtimes_{k= 1}^{K}\Xcal_k$. The expected value of a random variable (r.v.) $Z$ is denoted by $\EV[Z]$. The set of all Borel probability measures on $\Acal$ is denoted by $\Pcal(\Acal)$. Given a probability measure $\eta  \in \Pcal(\Acal)$, the mass on $a\in \Acal$ is denoted by $\eta(a)$.
In this paper, to characterize the distribution of mass of a population of mass $m>0$ over elements of a finite set $\Acal$ we use vectors $\mu \in X_{\Acal}:= \{ \nu \in  \Rnn^{|\Acal|} : \ones^\top \nu = m\}$. For the sake of clarity, by abuse of notation, the mass on $a\in \Acal$ is denoted by $\mu[a]$ and the mass on a subset $\Bcal\subseteq \Acal$ is denoted by $\mu[\Bcal]:= \sum_{a\in \Bcal}\mu[a]$.

%The unit simplex of dimension $n$ is the set denoted by $\Delta_n := \{x\in \Rnn^n  : \ones^\top x\leq 1 \}$.

%\revm{Notice that for all $c\in [C]$, the sets $\Scal^c$, $\Acal^c$, and $\Ucal_D^c$ are finite. Denote their carnality by $n^c:= |\Scal^c|$, $m^c := |\Acal^c|$, and $o^c:= |\Ucal_D^c|$, respectively, and let $n:= \sum_{c\in [C]} n^c$, $m:= \sum_{c\in [C]} m^c$, and $o:= \sum_{c\in [C]} o^c$. Notice that the concatenation of distribution maps $\mu_{\Scal \times \Acal}(t) \in X_{\Scal \times \Acal}$ and  $\mu(t) \in X$ can be completely characterized by vectors in $\R^{nm}$ and $\R^{no}$, respectively, whose entries are the image for each $c\in [C]$ of each element of $\Scal^c \times \Acal^c$ and $\Scal^c \times \Ucal^c_D$, respectively. Henceforth, by abuse of notation and whenever clear form the context, we use vector operations on  $\mu_{\Scal \times \Acal}(t) \in X_{\Scal \times \Acal}$ and  $\mu(t) \in X$.}

% !TeX spellcheck = en_US

%\vspace{-0.1cm}

\section{Model}\label{sec:model}

In this section, we present the model for a population of $N$ players and the mean field model approximation as $N \to \infty$.%, whose approximation properties are established throughout the paper. 

\vspace{-0.3cm}
\subsection{Finite-population Model}

The finite-population model is described by:
\begin{itemize}
	
	\item \emph{Population}: There are $N \in \N$ players which are spread across $C\in \N$ classes (also called subpopulations) with similar needs. We denote the class of a player $i\in [N]$ by $c^i$, which is time-invariant. The set of players that are in a class $c\in [C]$ is denoted by $\Ccal_c :=\{i\in [N]: c^i= c\}$.  The mass of players in a class $c\in \Ccal$ is denoted by $m^c := |\Ccal_c|/N$.
	
	\item \emph{Time}: Each player makes decisions in continuous time. Each player $i\in \Ccal_c$ is equipped with a Poisson clock with rate $\Rdc >0$ (which is equal to the rate of all other players in the same class). Each time the clock of a player rings, they take an action. We assume that clocks of different players are independent. The time of the $k$-th clock ring of a player $i\in [N]$ is characterized by a random variable (r.v.) $t^{i}_k$.
	
	\item \emph{States}: At each time $t \in [0,\infty)$, each player $i\in \Ccal_c$ has an individual state from a finite set of states $\Scal^c$ that evolves stochastically with their decisions, which is characterized by a r.v.\ $s^{i}(t)$. As a result, a realization of $s^{i}(t)$ has a piecewise-constant time evolution with discontinuities when the clock of the player rings. We also define $p^c := |\Scal^c|$ and $p:= \sum_{c\in [C]}p^c$.
	
	\item \emph{Actions}: The actions available to a player $i\in \Ccal_c$ in state $s\in \Scal^c$ are in the nonempty finite set $\Acal^c(s)$. We denote by $\Acal^c := \bigcup_{s\in \Scal} \Acal^c(s)$ the set of all actions available to a player of class $c$. The action that a player $i\in [N]$ would take at time $t$ if their clock were to ring is characterized by a r.v.\ $a^{i}(t)$. We also define $q^c := |\Acal^c|$ and $q:= \sum_{c\in [C]}q^c$.

	\item \emph{State transitions}: Upon an action of a player, their state evolves according to a Markov transition kernel $\phi^c: \Scal^c \times \Acal^c \to \Pcal(\Scal^c)$. We denote the distribution of the new state of a player in state $s \in \Scal^c$ that takes action $a\in \Acal^c(s)$ by $\phi^c(\cdot | s,a)$.
	
	\item \emph{State-action distribution}: The empirical joint state-action distribution of class $c\in [C]$ at time $t$ is characterized by a r.v.\ $\hat{\mu}^c_{\Scal \times \Acal}(t)$ with support in $X^c_{\Scal \times \Acal}:=\{\nu \in \Rnn^{p^cq^c} : \ones^\top\nu = m^c\}$. Recall that, by abuse of notation, $\hat{\mu}^c_{\Scal \times \Acal}[s,a](t)$ is the r.v.\ associated with the mass on $s\in \Scal^c$ and $a\in \Acal^c$ and it is given by $\hat{\mu}^c_{\Scal \times \Acal}[s,a](t) := \frac{1}{N}\sum_{i\in \Ccal_c} \delta_{s^{i}(t)}(s) \delta_{a^{i}(t)}(a)$. The concatenation of the empirical joint state-action distributions for all classes is denoted by $\hat{\mu}_{\Scal \times \Acal} = \col(\hat{\mu}^c_{\Scal \times \Acal}, c\in [C])$ with support in $X_{\Scal \times \Acal} := \bigtimes_{c\in [C]}X^c_{\Scal \times \Acal}$.
	
	\item \emph{Single-stage reward}: The single-stage reward of a player $i\in \Ccal_c$ is modeled by a real-valued function $r^c: \Scal^c \times \Acal^c \times X_{\Scal \times \Acal} \to \R$.  Specifically, the single-stage reward of a player in state $s\in \Scal^c$ that takes action $a\in \Acal^c(s)$ at time $t$ is $r^c(s,a,\hat{\mu}_{\Scal\times\Acal}(t))$. Notice that $N\sum_{c\in [C]: a\in \Acal^c}\Rdc\hat{\mu}^c_{\Scal\times\Acal}[\Scal^c,a](t)$ corresponds to the expected flow of players taking action $a$, which can be used to model a decreasing reward upon congestion of a resource, for instance. 
	
	\item \emph{Payoff}: The payoff of a player $i\in [N]$ is the long-time average reward which is given by
	\begin{equation*}
			J^i := \lim_{T\to \infty}\frac{1}{T}\EV\left[\sum_{k=1}^T r^{c^i}(s^i(t^i_k),a^i(t^i_k),\hat{\mu}_{\Scal \times \Acal}(t^i_k))\right].
	\end{equation*}

\end{itemize}

%	\vspace{-0.6cm}
	
\subsection{Policies}\label{sec:policies}

Given the information available to them, each player will take an action each time their clock rings. Loosely speaking, we call this map a \emph{policy}. Since we are particularly interested in an evolutionary analysis, where the players' knowledge is myopic, we consider the following \emph{information structure} on a policy of a player:
\begin{itemize}
	\item \emph{Oblivious}: The policy does not depend on any aggregate information about the distribution of the players' states. The dependence of the player decision on the state distribution is indirect through the rewards of each action. (This terminology was introduced in \cite{WeintraubBenkardEtAl2008} and policies with this property are studied in detail in similar games in \cite{AdlakhaJohariEtAl2015}).
	\item \emph{Markov}: The policy depends only on the individual state of a player at the time their clock rings.
	\item \emph{Stationary}: The policy is time-invariant, in the sense that when a player chooses a policy they plan to use it forever. %This property is not incompatible with an evolutionary framework where users revise their policies, it just means than when they 
\end{itemize}
A policy that is oblivious, Markov, and stationary can be characterized for a class $c\in [C]$ as a map $u:\Scal^c \to \Pcal(\Acal^c)$ from the state of the player when their clock rings to a randomization of actions. The set of such policies is denoted by $\Ucal^c$ and formally defined as
\begin{equation*}
	\Ucal^c := \{u:\Scal^c \to \Pcal(\Acal^c) \;|\; \mathrm{supp}(u(s))\subseteq \Acal^c(s), \forall s\in \Scal^c\}.
\end{equation*}
In general, the policies in $\Ucal^c$ are said to be \emph{randomized}, in the sense that they map a state to a randomized action. A particular case is a \emph{deterministic} policy, for which every state maps to a single action with probability one. The set of deterministic policies of a class $c\in [C]$ is denoted by $\Ucal^c_D\subset \Ucal^c$ and is formally defined as
\begin{equation*}
	\Ucal^c_D :=\{u \in \Ucal^c \;|\;  \forall s\in \Scal^c \;\exists a\in  \mathcal{A}^c(s) : \mathrm{supp}(u(s)) = \{a\} \}.
\end{equation*}
 We also define $n^c := |\Ucal_D^c|$ and $n:= \sum_{c\in [C]}n^c$.

We consider that at each time $t$ each player $i\in \Ccal_c$ uses a policy in $\Ucal_D^c$ that is characterized by a r.v.\ $u^{i}(t)$. In Section~\ref{sec:ev_model}, we introduce evolutionary dynamics to describe the time evolution of $u^{i}(t)$. Until then, we consider that the policy used by each player is constant in time, i.e., $u^{i}(t)$ is constant in time for all $i\in [N]$. The empirical joint state-policy distribution of class $c\in [C]$ is characterized by a r.v.\ $\hat{\mu}^c(t)$ with support in $X^c :=\{\nu \in \Rnn^{p^cn^c} : \ones^\top\nu = m^c\}$, which, by abuse of notation, is given by $\hat{\mu}^{c}[s,u](t) := \frac{1}{N}\sum_{i\in \Ccal_c}^{N} \delta_{s^{i}(t)}(s) \delta_{u^{i}(t)}(u)$. The concatenation of the empirical joint state-policy distributions for all classes is denoted by $\hat{\mu} = \col(\hat{\mu}^c, c\in [C])$ with support in $X:= \bigtimes_{c\in [C]}X^c$.

%\vspace{-0.1cm}

\subsection{Mean Field Model}

A mean field model considers a continuum of players. Interestingly, the assumption on independent Poisson clocks allows to characterize the evolution of the distribution of states and actions in the population with ordinary differential equations (ODE). At time $t$, we denote the joint state-action distribution of class $c\in [C]$ by $\mu_{\Scal \times \Acal}^c(t) \in X^c_{\Scal \times \Acal}$ and the joint state-policy distribution of class $c\in [C]$ by  $\mu^c(t) \in X^c$. The concatenation of the joint state-action and state-policy distributions for all classes is denoted by $\mu_{\Scal \times \Acal} = \col(\mu^c_{\Scal \times \Acal}, c\in [C]) \in X_{\Scal \times \Acal}$  and $\mu = \col(\mu^c, c\in [C]) \in X$, respectively. Intuitively, for a class $c\in [C]$, in an infinitesimal interval of time $\dint t$, the difference in the mass in state $s\in \Scal^c$: (i)~increases by the proportion of clock rings in other states that, after taking an action, end up in state $s$; and  (ii)~decreases by the proportion of clock rings in state $s$ that take an action and leave the state; i.e., for all $s\in \Scal^c$ and $u \in \Ucal^c_D$,
\begin{equation}\label{eq:dt_mu_u_S}
	\begin{split}
 \dint \mu^c[s,u] &= \!\!\sum_{s^\prime\in \Scal^c}   \sum_{a^\prime\in \Acal^c(s^\prime)}\!\! \!\!\! \Rdc \mu^c[s^\prime,u] \dint t\phi^c(s|s^\prime\!,a^\prime)u(a^\prime|s^\prime) \\&- \Rdc\mu^c[s,u]\dint t \sum_{s^\prime\in \Scal^c} \sum_{a\in \Acal^c(s)}\phi^c(s^\prime|s,a)u(a|s).
\end{split}
\end{equation}
When $\dint t \to 0$ this balance equation can be written for all $s\in \Scal^c$ and $u \in \Ucal_D^c$ as
\begin{equation}\label{eq:ODE_mu_u_S}
	\begin{split}
	\dot{\mu}^c[s,u] &=  \Rdc \sum_{s^\prime \in \Scal^c} \sum_{a^\prime \in \Acal^c(s^\prime)}\phi^c(s|s^\prime,a^\prime)u(a^\prime|s^\prime)\mu^c[s^\prime,u] \\&- \Rdc\mu^c[s,u],
	\end{split}
	\end{equation}
	since $\sum_{s^\prime \in \Scal^c} \phi^c(s^\prime|s,a)  = 1$. The joint state-action distribution of class $c\in [C]$ follows from the solution to \eqref{eq:ODE_mu_u_S} for all $s\in \Scal^c$ and all $a \in \Ucal^c_D$ as 
	\begin{equation}\label{eq:mu_SA_infinite}
		\mu^c_{\Scal\times \Acal}[s,a](t) = \sum\nolimits_{u\in \Ucal_D^c}\mu^c[s,u](t)u(a|s).
	\end{equation}
	Notice that, contrarily to the time evolution of $\hat{\mu}$, the time evolution of $\mu$ is deterministic and governed by the ODE \eqref{eq:ODE_mu_u_S}.  It follows from the Picard-Lindel\"{o}f Theorem \cite[Theorem~5.7]{Smirnov2002}, that a solution $\mu(t)$ to \eqref{eq:ODE_mu_u_S} exists and is unique. Using Kurtz's Theorem \cite[Theorem~2.1 in Chap.~11]{EthierKurtz1986} we can show that  approximates arbitrarily well the evolution of the empirical joint state-policy distribution $\hat{\mu}(t)$ as $N\to \infty$, as formalized in the following result.

%	$\hat{\mu}^c(t)$ and  $\hat{\mu}^c_{\Scal \times \Acal}(t)$ converge almost surely to $\mu^c(t)$ and $\mu^c_{\Scal \times \Acal}(t)$, respectively, as $N\to \infty$, i.e.,
	\begin{lemma}\label{lem:approx_ODE_mu_u_S}
		For any class $c\in [C]$, a solution to \eqref{eq:ODE_mu_u_S} with initial condition $\mu^c(0) \in X^c$ exists in $t\in [0,\infty)$, is unique, and is Lipschitz continuous w.r.t.\ $\mu^c(0)$. Furthermore, if $\lim_{N\to \infty}\hat{\mu}^c(0)= \mu^c(0)$ almost surely, then $\lim_{N\to \infty}\hat{\mu}^c(t)= \mu^c(t)$ and $\lim_{N\to \infty}\hat{\mu}^c_{\Scal \times \Acal}(t)= \mu^c_{\Scal \times \Acal}(t)$ almost surely for all $t\in [0,\infty)$.
	\end{lemma}
	\begin{proof}
		See Appendix~\ref{sec:proof_lem_approx_ODE_mu_u_S}.
	\end{proof}

%	\vspace{-0.3cm}

\subsection{Assumptions}\label{sec:model_ass}

We impose a mild global continuous differentiability assumption on the single-stage reward, which in turn implies global Lipschitz continuity (since the domain of interest is compact). Continuity is necessary for the existence of equilibria. Lipschitz continuity is necessary for existence and uniqueness of trajectories to the ODE model of the evolutionary dynamics. Existence of a domain extension and continuous differentiability are necessary for the existence and continuity of partial derivatives.\footnote{The assumption on the existence of a domain extension could be lifted for most of the results presented in this paper by taking derivatives only along directions tangent to  $X_{\Scal \times \Acal}$. However, since it is very mild, it is kept for the sake of simplicity and clarity of the results.}

\begin{assumption}\label{ass:cont}
		For all $c \in [C]$, all $s\in \Scal^c$, and all $a\in \Acal^c$ the single-stage reward function $r^c(s,a,\mu_{\Scal\times \Acal})$ admits a domain extension to $\Scal^c \times \Acal^c \times \Rnn^{pq}$; and the domain extension is continuously differentiable w.r.t.\ $\mu^{d}_{\Scal\times \Acal}[s^\prime,a^\prime]$ for all $d\in[C]$, all $s^\prime\in \Scal^{d}$, and all $a^\prime \in \Acal^{d}$ in $\Scal^d\!\times\! \Acal^d \!\times\! X_{\Scal \times \Acal}$.
\end{assumption}

Furthermore, the analysis of the evolutionary dynamics under average payoff  is significantly simpler under the following mild regularity assumption on the state transition kernel of deterministic policies, which is weaker than an irreducibly assumption. For a detailed overview of elementary Markov chain analysis tools used in this paper see \cite{Norris1997}.

\begin{assumption}\label{ass:markov_unique}
	For all $c\in [C]$ and all $u\in \Ucal^c_D$, the state transition Markov kernel $\phi^{c,u}$ associated with policy $u$, which admits a matrix representation $\phi^{c,u}_{ss^\prime} =\sum_{a^\prime \in \Acal(s^\prime)}\phi^{c,u}(s|s^\prime,a^\prime)u(a^\prime|s^\prime)$ for all $s,s^\prime \in \Scal^c$, contains one and only one recurrent communicating class.
\end{assumption}

Assumption~\ref{ass:markov_unique} is weaker than the assumption introduced in \cite[Assumption A1]{WiecekAltman2015}, which also assumes that the unique recurrent communicating class is the same for all policies. Under Assumption~\ref{ass:markov_unique}, the continuous-time Markov chain associated with each policy converges almost surely to a unique stationary state distribution as shown for completeness in the following result. It is an extension of \cite[Theorem~3.5.2]{Norris1997} and \cite[Theorem~3.8.1]{Norris1997}, which only hold under the stronger assumption that the Markov chain associated with each policy is irreducible.

\begin{lemma}\label{lem:markov_unique}
	Under Assumption~\ref{ass:markov_unique}, for any class $c\in [C]$ and any policy $u\in \Ucal^c$, the continuous-time state transition Markov chain associated with $u$, which is generated by $Q^{c,u} = \Rdc(\phi^{c,u}-I)$, admits a unique stationary state distribution denoted by $\eta^{c,u} \in \Pcal(\Scal)$. Furthermore, the state distribution of a player $i\in [N]$ using policy $u\in \Ucal^c$ converges almost surely to $\eta^{c,u}$ from any initial condition as $t\to\infty$.
\end{lemma}
\begin{proof}
	See Appendix~\ref{sec:proof_lem_markov_unique}.
\end{proof}

Under Assumption~\ref{ass:markov_unique}, by Lemma~\ref{lem:markov_unique}, the long-time average reward of using policy $u\in \Ucal^c$ does not depend on the initial state distribution, therefore when the state-action distribution $\mu_{\Scal \times \Acal} \in X_{\Scal\times \Acal}$ is constant it can be written as
\begin{equation}\label{eq:def_J}
	J^c(u, \mu_{\Scal\times\Acal})\! :=\! \!\sum_{s\in\Scal^c} \sum_{a\in \Acal^c(s)} \!\!\!\eta^{c,u}(s)u(a|s)r^c(s,a,\mu_{\Scal \times \Acal}).
\end{equation}
To be more precise, the single-stage reward is continuous by Assumption~\ref{ass:cont} and defined in a compact set, therefore bounded, and the state distribution of a player using policy $u\in \Ucal^c$ converges almost surely to $\eta^{c,u}$ by Lemma~\ref{lem:markov_unique}. Thus, for a fixed $\mu_{\Scal \times \Acal}$, one can apply the Dominated Convergence Theorem \cite[Theorem~9.1.2]{Rosenthal2006} to express the long-time average reward as \eqref{eq:def_J}.
% !TeX spellcheck = en_US

\vspace{-0.1cm}

\section{Equilibria}\label{sec:equilibria}

In this section, we study NE-like solution concepts for the class of mean field games under consideration. The usefulness of a solution concept is naturally its ability to predict the outcome of the game. Before proceeding, we distinguish between two fundamentally distinct notions of a policy of the population: \emph{behavioral} and \emph{mixed}. On the one hand, we say that the population follows a \emph{behavioral} policy if each player of the same class uses the same randomization of actions for each state during the game, i.e., the same single policy in $\Ucal^c$ is chosen by every player $i\in \Ccal_c$. On the other hand, we say that a population follows a \emph{mixed} policy if each player randomizes over deterministic policies ex ante, i.e., at the start of the game each player $i\in \Ccal_c$ chooses a deterministic strategy in $\Ucal^c_D$ and sticks with it forever.\footnote{A qualitatively analogous distinction is made in the context of extensive games (for more information see \cite[Parts II and III]{OsborneRubinstein1994}) from which we borrowed the terminology and qualitative intuition. Even though under an assumption of perfect recall these notions are equivalent in the context of extensive games \cite[Proposition~99.2]{OsborneRubinstein1994}, that is not the case for the class of mean field games at hand.}

The \emph{behavioral} approach of defining a solution concept for finite-state mean field games has been given almost exclusive attention in the literature. In this section, first, we argue that a \emph{behavioral} solution concept has some deficiencies and it may not be a qualitatively or quantitatively good prediction of the outcome of the game. Second, we proceed by proposing a novel \emph{mixed} solution concept, which is arguably more natural in this context and appears to have not been studied yet in continuous-time finite-state mean field games. Third, we establish theoretical foundations for the novel mixed solution concept, namely existence and approximation properties w.r.t.\ the analogous finite-population dynamic game. Furthermore, Section~\ref{sec:ev_model} reveals that a mixed solution concept is instrumental for an intuitive evolutionary interpretation and notion of evolutionary stability, which is not obtainable taking a behavioral approach. In Section \ref{sec:MAC_equilibria}, the comparison between both solution concepts is illustrated for a simple game.

The literature on continuous-time finite-state stochastic dynamic games of many players (and similar classes of mean field games) focuses exclusively on the concept of a behavioral stationary Nash equilibrium (BSNE) as a solution concept (e.g.~\cite{AdlakhaJohariEtAl2015,WiecekAltman2015,Wiecek2020,ElokdaBolognaniEtAl2024}), which is informally defined as:
\begin{mdframed}[style=callout]
			A \emph{behavioral stationary Nash equilibrium} (BSNE) is an equilibrium condition whereby all players of the same class $c\in [C]$ use the same (randomized) policy $u \in \Ucal^c$ (the population uses a behavioral policy) such that: (i)~the resulting state distribution is stationary; and (ii)~no player can unilaterally deviate from $u$ to another policy $v\in \Ucal^c$ to increase their payoff. 
\end{mdframed}
%Intuitively, the population plays a behavioral strategy in the sense defined above, whereby no agent has an incentive to change policy and that leads to the aggregate state stationarity. 
Contrarily, we informally define a mixed stationary Nash equilibrium (MSNE) as:
\begin{mdframed}[style=callout]
	A \emph{mixed stationary Nash equilibrium} (MSNE) is an equilibrium condition whereby each player of a class $c\in[C]$ uses a deterministic policy $u \in \Ucal^c_D$ (the population uses a mixed policy) such that: (i)~the resulting state distribution is stationary; and (ii)~no player can unilaterally deviate from $u$ to another policy $v \in \Ucal_D^c$ to increase their payoff. 
\end{mdframed}

The behavioral and mixed solution concepts defined above have fundamentally different natures in two key aspects. First, a behavioral solution concept relies on the notion of randomization of actions by a single player. Historically, there has been considerable debate on whether such modeling approach is meaningful in real-life applications. The interested reader is referred to the interesting discussion on this topic in \cite[Chap.~3.2]{OsborneRubinstein1994}, where curiously the two authors of the book have distinct views. Notably, the mixed solution concept does not require this notion of randomization of actions, instead each player chooses a deterministic action.\footnote{The use of the term ``mixed solution concept'' should not be confused with a ``mixed strategy'' in normal-form games. Even thought this terminology can lead to confusion, we stick to it for the sake of consistency with related literature, e.g., \cite{AltmanHayel2010}.} Second, even if for a particular application the randomization of actions is understood to be realistic, \emph{there is no physically meaningful reason for the policies adopted by every player in each class to be the same}, as it is assumed in the definition of the behavioral solution concept. Finally, a BSNE is, in general, easier to compute numerically. In spite of that, the physical meaningfulness is the priority for the evolutionary analysis in this paper. This point will be discussed further in the conclusion section.

\vspace{-0.2cm}

\subsection{Definition of BSNE and MSNE}

In this section, we present the formal definitions of BSNE and MSNE.

\begin{definition}[BSNE]\label{def:BSNE}
	For each class $c\in [C]$, consider a policy $u_c\in \Ucal^c$. The collection $(u_c,\eta^{c,u_c})_{c\in [C]}$ is said to be a BSNE in the average payoff mean field game if
	\begin{equation*}
		J^c(u_c,\mu_{\Scal\times\Acal})  \geq J^c(v,\mu_{\Scal\times\Acal}), \quad \forall c\in [C]\; \forall v\in \Ucal^c
	\end{equation*}
	where $\mu_{\Scal\times\Acal} \in X_{\Scal \times \Acal}$ is characterized by $\mu^c_{\Scal\times\Acal}[s,a] = m^c \eta^{c,u_c}(s) u_c(a|s)\;  \forall s\in \Scal^c \; \forall a\in \Acal^c$ and $\eta^{c,u_c} \in \Pcal(\Scal^c)$ is the stationary state distribution of the continuous-time state transition Markov chain associated with $u_c$, which is unique by Lemma~\ref{lem:markov_unique} under Assumption~\ref{ass:markov_unique}. \hfill$\triangle$
\end{definition}

Since the MSNE relies only on a finite number of policies for each class, define for all $c\in [C]$ a payoff map $F^c: X\to \R^{n^c}$ as 
\begin{equation}\label{eq:def_F}
	F^c(\mu) = \col(J^c(u,\mu_{\Scal \times \Acal}), u\in \Ucal_D^c),
\end{equation}
where $\mu_{\Scal \times \Acal}$ is written as a function of $\mu$ resorting to \eqref{eq:mu_SA_infinite}. For the sake of clarity, by abuse of notation, we denote the component associated with policy $u\in \Ucal_D^c$ by $F^c_u(\mu)$. We also write the concatenation of the payoff maps of all classes as a payoff map $F^c: X\to \R^{n}$ given by $F(\mu) = \col(F^c(\mu), c\in [C])$. Notice that, when restricted to deterministic policies, the dynamic mean field game can be fully characterized by the pair $(F,\phi)$, where $\phi = (\phi^{c,u})_{c\in [C],u\in \Ucal_D^c}$.

\vspace{0.2cm}

\begin{definition}[MSNE]\label{def:MSNE}
	A joint state-policy distribution $\mu \in X$ is said to be a MSNE in the average payoff mean field game, denoted by $\mu \in \MSNE(F,\phi)$, if for all $c\in [C]$ and all $u\in \Ucal^c_D$
	\begin{align}
		&\mu^c[\Scal^c,u] >0 \implies   F^c_u(\mu)  \geq F^c_v(\mu), \quad \forall v\in \Ucal^c_D \label{eq:MSNE_u}\\
		&\mu^c[s,u] = \eta^{c,u}(s)\mu[\Scal^c,u]  \quad \forall s\in \Scal^c.   \label{eq:MSNE_s}
	\end{align}%
	where $\mu^c[\Scal^c,u] = \sum_{s\in \Scal^c}\mu^c[s,u]$ for all $c\in [C]$ and all $u\in \Ucal^c_D$, and $\eta^{c,u} \in \Pcal(\Scal^c)$ is the stationary state distribution of the continuous-time state transition Markov chain associated with $u$, which is unique by Lemma~\ref{lem:markov_unique} under Assumption~\ref{ass:markov_unique}. \hfill$\triangle$
\end{definition}

\begin{mdframed}[style=callout]
	It is interesting to note the particularly simple and intuitive definition of the MSNE. It is an equilibrium condition where each individual player uses a possibly different deterministic policy in steady-state and has no incentive to switch from their policy to any other deterministic policy. This intuitive definition is instrumental to define evolutionary dynamics and to study the evolutionary stability of equilibria in Part~II of this work.
\end{mdframed}

	\vspace{-0.4cm}

%\subsection{MSNE Properties: Existence}
\subsection{Existence}

In this section, we establish the existence of at least one MSNE. Before that, we introduce the notion of \emph{steady-state game}, which is a payoff map that particularizes $F$ when the state dynamics are stationary. 

\begin{definition}
	For each class $c\in [C]$, define a payoff map $\Fcal^c: X^c_{\Ucal_D} \to \R^n$ as $\Fcal^c(x) = F^c(\bar{\mu}(x))$. Here $\bar{\mu}(x) \in X$ is the stationary state-policy distribution associated with a marginal policy distribution $x^c\in X^c_{\Ucal_D}:= \{\nu \in \Rnn^{n^c}: \ones^\top\nu = m^c\}$, which is characterized by $\bar{\mu}^c(x)[s,u] = x^c[u]\eta^{c,u}(s)$ for all $c\in [C]$, all $s\in \Scal^c$, and all $u\in \Ucal_D^c$. We also define $X_{\Ucal_D}:= \bigtimes_{c\in[C]} X^c_{\Ucal_D}$. The \emph{steady-state game} is a payoff map $\Fcal: X_{\Ucal_D} \to \R^n$ characterized by $\Fcal(x)  = \col(\Fcal^c(x), c\in [C])$. \hfill $\triangle$
\end{definition}

%\begin{equation*}
%\quad \text{with} \bar{\mu}(x) \in \quad \bar{\mu}(x)[s,u] = x(u)\eta^u(s).}
%\end{equation*}
Notice that we can only define such a steady-state game game due to Lemma~\ref{lem:markov_unique} under Assumption~\ref{ass:markov_unique}. Interestingly, the \emph{steady-state game} admits a standard notion of NE, which is defined in what follows.

\begin{definition}\label{def:NEss}
		A policy distribution $x\in X_{\Ucal_D}$ is said to be a NE of the steady-state game $\Fcal$,  denoted by $x\in \NE(\Fcal)$, if for all $c\in [C]$ and all $u\in \Ucal^c_D$ \;$x^c[u]>0 \implies \Fcal^c_u(x)\geq \Fcal^c_v(x)\; \forall v\in \Ucal^c_D$. \hfill$\triangle$
\end{definition}

One can establish an equivalence between $\NE(\Fcal)$ and $\MSNE(F,\phi)$, as shown in the following lemma.

\begin{lemma}\label{lem:NEss_iff_MSNE} 
	Under Assumption~\ref{ass:markov_unique}, consider $x\in X_{\Ucal_D}$ and $\mu\in X$ defined as $\mu^c[s,u] = \eta^{c,u}(s)x^c[u], \; \forall c\in [C]\; \forall s \in \Scal^c,\forall u\in \Ucal_D^c$. Then $x\in \NE(\Fcal)  \iff \mu \in \MSNE(F,\phi)$.
\end{lemma}
\begin{proof}
	Both directions of the equivalence follow directly from comparing Definitions~\ref{def:MSNE} and~\ref{def:NEss}.
\end{proof}

This means that static properties of the dynamic game, like the properties of the MSNE, can be studied resorting to the analysis of the \emph{steady-state game} using simple and known static results. Naturally, dynamic properties such as evolutionary stability cannot leverage this relation.

The following result establishes the existence of a MSNE. Results on the existence of at least one BSNE in this setting can be obtained using similar arguments as in the existence results in \cite{WiecekAltman2015,Wiecek2020}. 

% It relies on writing the MSNE as a fixed point of a set-valued map and then using Kakutani's fixed point theorem.
\begin{theorem}\label{th:existence_MSNE}
	Under Assumptions~\ref{ass:cont} and~\ref{ass:markov_unique}, $(F,\phi)$ admits at least one MSNE.
\end{theorem}
\begin{proof}
	Under Assumption~\ref{ass:cont}, the steady-state payoff map $\Fcal$ is continuous. Therefore, it follows from a well-known result~\cite[Proposition~33.1]{OsborneRubinstein1994} that since $n$ is finite, $\NE(\Fcal)$ is nonempty. Under Assumption~\ref{ass:markov_unique}, by Lemma~\ref{lem:NEss_iff_MSNE} one concludes that $\MSNE(F,\phi)$ is nonempty.
\end{proof}

%\subsection{MSNE Properties: Uniqueness under Congestion Game Payoff Structure}
\subsection{Uniqueness under Congestion Game Payoff Structure}

An interesting particular payoff structure can arise, whose steady-state game is analogous to the well-known class of congestion games~\cite{Sandholm2001}. This setting is explored in the following example.

\begin{example}\label{eg:congestion}
	Assume that there is a finite collection of resources $\Rcal$ (e.g., road links in an urban network). Let the action rate be the same for every class, i.e., $\Rd = \Rd^1 = \Rd^2, \cdots = \Rd^C$. Every action $a\in \Acal^c$ is a subset of the resources, i.e., $a \subseteq \Rcal$ (e.g., each action is a path that uses a subset of the road links). One can also denote the set of actions of class $c\in [C]$ that use a resource $r\in \Rcal$ by $\Acal^c_{\Rcal}(r) \subseteq \Acal$. Each resource $r\in \Rcal$ has a reward function $w_r:\Rnn\to \R$  (e.g., the symmetric of the travel time on road link $r$), which is a function of the flow of players using resource $r$, denoted by $\sigma_r =  \Rd\sum_{c\in [C]}\sum_{s\in \Scal^c}\sum_{a\in \Acal^c_\Rcal(r)} \mu^c_{\Scal \times \Acal}[s,a]$. Assume that the single-stage reward of an action is given by the sum of the resources' payoffs that it uses, i.e., for a class $c \in [C]$, $r^c(s,a,\mu_{\Scal\times \Acal}) = \sum_{r\in a}w_r(\sigma_r)$. Henceforth, we refer to this payoff structure as a \emph{congestion game payoff structure}. Additionally, we refer to a \emph{decreasing (nonincreasing) rewards congestion game payoff structure} when  the resources' reward functions $w_r$ are strictly decreasing (nonincreasing) for all $r\in \Rcal$. 
\end{example}

Notice that according to a \emph{congestion game payoff structure} in Example~\ref{eg:congestion}, the single-stage reward depends only on the action and marginal action distribution of the mean field.  In this case, the state can only shape the average payoff of a player $i \in \Ccal_c$ through the admissible set of actions for each state $\Acal^c(s)$. Notice that the token economy setting in Example~\ref{eg:token} follows this structure. In Section~\ref{sec:MAC_equilibria}, we analyze a real-life example of a medium access game between mobile terminals competing for a common wireless channel which does not satisfy this payoff structure.
Nevertheless, under the congestion game payoff structure, the steady-state game has very strong properties, which are described in the following result.

\begin{lemma}\label{lem:congestion_ss_is_potential}
	Under a congestion game payoff structure (see Example~\ref{eg:congestion}) and Assumptions~\ref{ass:cont} and~\ref{ass:markov_unique}, the steady state-game is a full potential game, i.e., there exists a continuously differentiable function $U:\Rnn^n\to \R$ such that $\Fcal = \nabla U$. Furthermore, under a nonincreasing rewards congestion game payoff structure, $\NE(\Fcal)$ is compact and convex and, under a decreasing rewards congestion game payoff structure, the equilibrium resource flows $\sigma_r$ with $r\in \Rcal$ are unique.
\end{lemma}
\begin{proof}
	Define $U(x) = (1/\Rd)\sum_{r\in \Rcal}\int_0^{\sigma_r(x)} w_r(z) \dint z$, where $\sigma_r(x) = \Rd\sum_{c\in [C]}\sum_{s\in\Scal^c} \sum_{u\in \Ucal_D^c: u(s) \in \Acal^c_{\Rcal}(r)}\eta^{c,u}(s)x^c[u]$ is the steady-state flow of players using resource $r\in \Rcal$, which is well-defined and unique by Lemma~\ref{lem:markov_unique} under Assumption~\ref{ass:markov_unique}. The first statement follows from the fact that for all $c\in [C]$ and all $u^c\in \Ucal_D$
	\begin{equation*}
		\frac{\partial U(x)}{\partial x^c[u]} =  \sum_{c\in [C]}\sum_{s\in \Scal^c}\eta^{c,u}(s)\sum_{r\in u(s)}w_r(\sigma_r) = \Fcal^c_u(x),
	\end{equation*}
	which is continuous by Assumption~\ref{ass:cont}.
	If $w_r$ are nonincreasing (decreasing) then $U$ is concave (strictly concave w.r.t.\ $\sigma_r$), thus the equilibria analysis reduces to analysis in \cite[Proposition~3.1]{Sandholm2001} and \cite[Exercise~3.1.5]{Sandholm2010}, which shows the second statement.
\end{proof}

Together with the relation between $\NE(\Fcal)$ and $\MSNE(F,\phi)$ in Lemma~\ref{lem:NEss_iff_MSNE}, Lemma~\ref{lem:congestion_ss_is_potential} leads immediately to a uniqueness result of the MSNE of the mean field game.

\begin{corollary}
	Under a nonincreasing rewards congestion game payoff structure (see Example~\ref{eg:congestion}), $\MSNE(F,\phi)$ is compact and convex. Furthermore, under a decreasing rewards congestion game payoff structure, the equilibrium resource flows $\sigma_r$ with $r\in \Rcal$ are unique.
\end{corollary}

\vspace{-0.2cm}

%\subsection{MSNE Properties: Approximation}\label{sec:approx_MSNE}
\subsection{Approximation}\label{sec:approx_MSNE}
%Approximation of Equilibria of Finite-population Game

In this section, we define a concept of equilibrium for the finite-population game that is analogous to the MSNE. Then, we establish that as $N\to\infty$ the MSNE in the mean field game is a good approximation. Analogous approximation results for the BSNE can be derived using similar arguments as in \cite{WiecekAltman2015}. Consider a collection of players' policies $\{u^i\}_{i\in [N]}$ and denote the long-time average reward of player $i\in [N]$ in the finite-population setting by
\begin{equation*}
	\begin{split}
		&J^{i,N}( u^1, u^2, \ldots,u^i, \ldots, u^N) \\
	 = & \lim_{T\to \infty}\frac{1}{T}\EV\left[\sum_{k=1}^T r^{c^i}(s^i(t^i_k),a^i(t^i_k),\hat{\mu}_{\Scal \times \Acal}(t^i_k))\right].
	\end{split}
\end{equation*}%
The definition of a weak MSNE in the average payoff finite-population game is as follows.

%\begin{definition}\label{def:MSNE_finite}
%	The collection  of players' policies  $\{u^i\}_{i\in [N]}$ is said to be a weak MSNE in the average payoff finite-population game if for all $i\in [N]$ and all 	$v^i\in\Ucal^{c^i}_D$
%	\begin{equation}\label{eq:MSNE_finite_u}
%		J^{i,N}( u^1, \ldots,u^i, \ldots, u^N) \geq J^{i,N}(u^1, \ldots,v^i, \ldots, u^N),
%	\end{equation}%
% and it is said to be a weak $\epsilon$-MSNE if
% \begin{equation*}
% 	J^{i,N}( u^1, \ldots,u^i, \ldots, u^N) \geq J^{i,N}(u^1, \ldots,v^i, \ldots, u^N) -\epsilon
% \end{equation*}%
% for some $\epsilon>0$.\hfill$\triangle$
%\end{definition}

\begin{definition}\label{def:MSNE_finite}
	The collection  of players' policies  $\{u^i\}_{i\in [N]}$ is said to be a weak  $\epsilon$-MSNE for some $\epsilon>0$ in the average payoff finite-population game if for all $i\in [N]$ and all 	$v^i\in\Ucal^{c^i}_D$
	\begin{equation*}
		J^{i,N}\!( u^1\!, \ldots,u^i\!, \ldots, u^N) \! \geq \! J^{i,N}\!(u^1\!, \ldots,v^i\!, \ldots, u^N) \!-\!\epsilon. \qquad  \triangle 
	\end{equation*}
\end{definition}

Intuitively, the collection  $\{u^i\}_{i\in [N]}$ is a weak MSNE of the finite-population game if each player cannot switch to another deterministic policy to obtain a better outcome. Crucially, the following result establishes that a MSNE in the mean field game approximates arbitrarily well, for large enough $N$, a weak MSNE in the finite-population game. 

\begin{theorem}\label{th:approx_MSNE}
	Let $\mu$ be a MSNE in the average payoff mean field game according to Definition~\ref{def:MSNE}.  Then, for any $\epsilon >0$ there is $N_\epsilon \in \N$ such that for any $N>N_\epsilon$, any collection of policies $\{u^i\}_{i\in [N]}$ of a finite population of $N$ players that satisfies
	\begin{equation}\label{eq:cond_N_dist_pol}
		\bigg|\frac{1}{N}\!\sum\nolimits_{i\in \Ccal_c}\!\!\delta_{u^i}(u) - \mu^c[\Scal^c,u]\bigg|< \frac{1}{N}, \quad \!\!\forall c\in [C]\; \forall u\in \Ucal^c_D
	\end{equation}
	is a weak $\epsilon$-MSNE in the finite-population game.
\end{theorem}
\begin{proof}
	See Appendix~\ref{sec:proof_th_approx_MSNE}.
\end{proof}

% !TeX spellcheck = en_US
\section{Evolutionary Dynamical Model}\label{sec:ev_model}

In Section~\ref{sec:equilibria}, we study solution concepts that predict the outcome of strategic interactions between players based on a game-theoretical notion of equilibrium. In this section, we turn to the individual behavior of the players playing them. Specifically, we propose an \emph{evolutionary dynamical model} where players occasionally \emph{revise} their choices.

%First, we obtain the mean field evolutionary model described by an ODE and show that it approximates arbitrarily well the finite-population evolutionary dynamics as $N\to \infty$. Second, we study the relation between a rest point of the evolutionary dynamics and the equilibrium solution concepts in Section~\ref{sec:equilibria}.

\subsection{Individual Evolutionary Dynamics}\label{sec:ev_dynamics}
Evolutionary models are very well studied for static games (also known as population games) \cite{Sandholm2010}, which are games where players do not possess an individual state that influences their reward and action space. The foundations of the evolutionary model for dynamic games presented in this section rely on that literature. Individual players revise their choices individually, which is expressive of \emph{inertia} and \emph{myopia} properties of behavior seen in real-life (and not collaboratively as an agreement of a population, which is rare in a large population). As a result, the evolutionary model proposed in this section naturally relies on modeling the evolution of the decision of individual players, specifically of the (deterministic) policies that they use. By abuse of notation, for a class $c\in [C]$, we denote the vector of the mass on each policy as $\hat{\mu}^c[\Scal^c,\cdot](t) := \col(\hat{\mu}^c[\Scal^c,u](t),u\in \Ucal_D) \in X^c_{\Ucal_D}$. Henceforth, the time dependence is oftentimes dropped for conciseness. The evolutionary model is described by: 
\begin{itemize}
	\item \emph{Time}: Each player makes revisions in continuous time. Each player $i\in \Ccal_c$ is equipped with a Poisson clock with rate $\Rrc >0$ (which is equal to the rate of all other players in the same class). Each time the clock of a player rings, they have the opportunity to revise the policy that they are currently using. We assume that action and revision clocks of all players are independent.
	
	\item \emph{Policy transitions}: Upon a revision opportunity of a player, their policy choice evolves according to a \emph{revision protocol}. A revision protocol of a class $c\in [C]$ is a map $\rho^c : \R^{n^c} \times X^c_{\Ucal_D} \to \Rnn^{n^c\times n^c}$, where the component associated with the pair $(u,v) \in \Ucal^c_D\times \Ucal^c_D$ is denoted, by abuse of notation, by $\rho^c_{uv}$. Specifically, a player using policy $u\in \Ucal^c_D$ switches to policy $v\in \Ucal^c_D$ with a switch rate $\rho^c_{uv}(F^c(\hat{\mu}),\hat{\mu}^c[\Scal^c,\cdot])$, where $F^c(\hat{\mu})$ is defined in \eqref{eq:def_F} and the policy ordering of $F^c(\hat{\mu})$ and of $\hat{\mu}^c[\Scal^c,\cdot]$ is consistent.

%		     a vector of the average payoff for each policy $u_1, \ldots,u_{|\Ucal_D|} \in \Ucal_D$, i.e., 
%	\begin{equation}\label{eq:def_Fs}
%		F(\hat{\mu}) := \col\left(J(u_1,\hat{\mu}_{\Scal\times \Acal}), \ldots, J(u_{|\Ucal_D|},\hat{\mu}_{\Scal\times \Acal})\right),
%	\end{equation}
%	where the policy ordering is consistent with the definition of $\hat{\mu}[\Scal,\cdot]$, and $\hat{\mu}_{\Scal\times\Acal}[s,a] = \sum_{u\in \Ucal_D}\hat{\mu}[s,u]u(a|s)\;  \forall s\in \Scal \; \forall a\in \Acal$. By abuse of notation we also write $F_u(\hat{\mu}) = J(u,\hat{\mu}_{\Scal\times \Acal})$.
\end{itemize}

% I'm not sure wether we need this still:
%- For any state $s\in \Scal$, we denote the vector of the mass on each policy as $\hat{\mu}[s,\cdot](t) := \col(\hat{\mu}[s,u](t),u\in \Ucal_D) \in \Delta_{|\Ucal_D|}$.

Intuitively, if a player $i\in \Ccal_c$ using policy $u\in \Ucal_D^c$ receives a revision opportunity, they switch to a policy $v\in \Ucal^c_D$ with probability $\rho^c_{uv}(F^c(\hat{\mu}),\hat{\mu}^c[\Scal^c,\cdot])/\Rrc$, and they continue to use the same policy with probability $1-\sum_{v\neq u}\rho^c_{uv}(F^c(\hat{\mu}),\hat{\mu}^c[\Scal^c,\cdot])/\Rrc$. We make an assumption to ensure that the aforementioned switching probabilities are well defined and continuous as follows.
\begin{assumption}\label{ass:rev_protocol}
	For all $c\in [C]$, the revision protocol $\rho^c$ is Lipschitz continuous and for all $u\in \Ucal^c_D$
	\begin{equation*}
		\Rrc \geq \sup_{\substack{\mu \in X}} \sum_{v\in \Ucal^c_D \setminus \{u\}} \rho^c_{uv}(F^c(\mu),\mu^c[\Scal^c,\cdot]).
	\end{equation*}
\end{assumption}

The literature on evolutionary decision dynamics identifies physically meaningful classes of revision protocols. In this paper, we restrict our attention to \emph{deterministic}\footnote{Revisions protocols are said to be deterministic if they generate unique solutions for the evolution of the aggregate decisions. Lipschitz continuity of $\rho^c$ in Assumption~\ref{ass:rev_protocol} ensures that $\rho^c$ is deterministic.} revision protocols, whose main classes are defined below.

%\begin{definition}[Imitative {\cite[Chap.~5.4]{Sandholm2010}}]\label{def:imitative}
%	Consider a Lipschitz continuous conditional imitation rate map $r^c :\R^{n_c} \times X_{\Ucal_D^c}\to \Rnn^{n^c\times n^c}$ with monotone net conditional imitation rates, i.e., $F^c_v \geq F^c_u \iff r^c_{kv}(F^c,\sigma) -   r^c_{vk}(F^c,\sigma)  \geq  r^c_{ku}(F^c,\sigma) -   r^c_{uk}(F^c,\sigma), \forall F^c \in \R^{n^c}\;\forall \sigma \in  X_{\Ucal_D^c}\;\forall u,v,k \in \Ucal^c_D$. Then, the revision protocol $\rho^c$ defined as $\rho^c_{uv}(F^c,\sigma) = r^c_{uv}(F^c,\sigma)\sigma_v/m^c$ is called an \emph{imitative} revision protocol. \hfill$\triangle$
%\end{definition}

\begin{definition}[Imitative {\cite[Chap.~5.4]{Sandholm2010}}]\label{def:imitative}
	Consider a revision protocol $\rho^c$ defined as $\rho^c_{uv}(F^c,\sigma) = r^c_{uv}(F^c,\sigma)\sigma_v/m^c$, where $r^c :\R^{n_c} \times X_{\Ucal_D^c}\to \Rnn^{n^c\times n^c}$ is a Lipschitz continuous conditional imitation rate map  with monotone net conditional imitation rates, i.e., $F^c_v \geq F^c_u \iff r^c_{kv}(F^c,\sigma) -   r^c_{vk}(F^c,\sigma)  \geq  r^c_{ku}(F^c,\sigma) -   r^c_{uk}(F^c,\sigma), \forall F^c \in \R^{n^c}\;\forall \sigma \in  X_{\Ucal_D^c}\;\forall u,v,k \in \Ucal^c_D$. Then $\rho^c$ is said to be an \emph{imitative} revision protocol. \hfill$\triangle$
\end{definition}

\begin{definition}[Imitative via comparison]\label{def:imitative_cmp}
	Consider an imitative revision protocol $\rho^c$  according to Definition~\ref{def:imitative}  characterized by $\rho^c_{uv}(F^c,\sigma) = r^c_{uv}(F^c,\sigma)\sigma_v/m^c$. The protocol $\rho^c$ is called an \emph{imitative via comparison} protocol if the imitation rates are sign-preserving, i.e., $\sign(r^c_{uv}(F^c,\sigma)) = \sign(\max(0,F^c_v-F^c_u)), \forall F^c \in \R^{n^c} \; \forall \sigma \in  X_{\Ucal_D^c}\; \forall u,v \in \Ucal^c_D$. \hfill$\triangle$
\end{definition}

%\begin{definition}[{Excess payoff \cite[Chap.~5.5]{Sandholm2010}}]\label{def:excess}
%	Consider a Lipschitz continuous rate map $\tau^c :\R^{n^c} \to \Rnn^{n^c}$ that satisfies acuteness, i.e.,
%	$\hat{F}^c \in \R^{n^c} \setminus \Rnp^{n^c} \!\implies \! \tau^c(\hat{F}^c)^\top\hat{F}^c > 0$.
%	Then, the revision protocol $\rho^c$ defined as $\rho^c_{uv}(F^c,\sigma) = \tau_{v}^c(\hat{F}^c)$, where $\hat{F}^c := F^c - \ones {F^c}^\top\sigma/m^c$ represents the excess payoff vector, is called an \emph{excess payoff} revision protocol. Furthermore, $\rho^c$ is said to be a \emph{separable excess payoff} revision protocol if $\tau_{v}^c(\hat{F}^c) \equiv \tau_{v}^c(\hat{F}_v^c)$. \hfill$\triangle$
%\end{definition}

\begin{definition}[{Excess payoff \cite[Chap.~5.5]{Sandholm2010}}]\label{def:excess}
	Consider a revision protocol $\rho^c$ defined as $\rho^c_{uv}(F^c,\sigma) = \tau_{v}^c(\hat{F}^c)$,  where $\hat{F}^c := F^c - \ones {F^c}^\top\sigma/m^c$ represents the excess payoff vector and $\tau^c :\R^{n^c} \to \Rnn^{n^c}$ is a Lipschitz continuous rate map that satisfies acuteness, i.e.,
	$\hat{F}^c \in \R^{n^c} \setminus \Rnp^{n^c} \!\implies \! \tau^c(\hat{F}^c)^\top\hat{F}^c > 0$. Then $\rho^c$ is called an \emph{excess payoff} revision protocol. Furthermore, $\rho^c$ is said to be a \emph{separable excess payoff} revision protocol if $\tau_{v}^c(\hat{F}^c) \equiv \tau_{v}^c(\hat{F}_v^c)$. \hfill$\triangle$
\end{definition}

%\begin{definition}[{Pairwise comparison \cite[Chap.~5.6]{Sandholm2010}}]\label{def:pairwise_cmp}
%	Consider a Lipschitz continuous rate map $\tau :\R^{n^c} \to \Rnn^{n^c}$ that is sign-preserving, i.e., $\sign(\tau^c_{uv}(F)) = \sign(\max(0,F^c_v-F^c_u)), \forall F^c
%	\in \R^{n^c} \; \forall u,v \in \Ucal^c_D$.
%	Then, the revision protocol $\rho^c$ defined as $\rho^c_{uv}(F^c,\sigma) = \tau^c_{uv}(F^c)$ is called a \emph{pairwise comparison} revision protocol.  Furthermore, if $\rho_{uv}^c(F^c,\sigma) = \phi^c_v(F^c_v-F^c_u)$ for some functions $\phi^c_v:\R \to \Rnn$, then $\rho^c$ is said to be an \emph{impartial pairwise comparison} revision protocol. \hfill$\triangle$
%\end{definition}

\begin{definition}[{Pairwise comparison \cite[Chap.~5.6]{Sandholm2010}}]\label{def:pairwise_cmp}
	Consider a revision protocol $\rho^c$ defined as $\rho^c_{uv}(F^c,\sigma) = \tau^c_{uv}(F^c)$, where $\tau :\R^{n^c} \to \Rnn^{n^c}$ is a Lipschitz continuous rate map  that is sign-preserving, i.e., $\sign(\tau^c_{uv}(F)) = \sign(\max(0,F^c_v-F^c_u)), \forall F^c \in \R^{n^c} \; \forall u,v \in \Ucal^c_D$. Then $\rho^c$  is called a \emph{pairwise comparison} revision protocol. Furthermore, if $\rho_{uv}^c(F^c,\sigma) = \phi^c_v(F^c_v-F^c_u)$ for some functions $\phi^c_v:\R \to \Rnn$, then $\rho^c$ is said to be an \emph{impartial pairwise comparison} revision protocol. \hfill$\triangle$
\end{definition}

These families of protocols follow from an intuitive interpretation of meaningful decision dynamics:
\begin{enumerate}[(a)]
	\item \emph{Imitative}: When the revision clock of a player rings and they have the opportunity to revise their policy, they choose a random player from their class. If the player is using policy $u\in \Ucal^c_D$ and the randomly chosen player is using policy $v\in \Ucal^c_D$, then the player will imitate the policy of the random player with a probability that is proportional to the imitation rate $r^c_{uv}$, i.e., $r^c_{uv}/\Rrc$. In turn, the imitation rate is such that if policy $v\in \Ucal^c_D$ has a higher payoff than policy $u\in \Ucal^c_D$, then the net imitation rate from any strategy $k\in \Ucal^c_D$ to $v$  has to be greater of equal than the rate from $k$ to $u$, which is portrayed in Definition~\ref{def:imitative}.
	\item  \emph{Excess payoff}:  Assume the players have access to the average payoff of each class (e.g., through information aggregation of a central planner). The decisions of players of class $c$ are based on comparing the payoff of the policies in $\Ucal_D^c$, i.e., $F^c(\hat{\mu})$, with the average payoff of the population, i.e., $F^c(\hat{\mu})^\top\hat{\mu}^c[\Scal^c,\cdot]/m^c$. The weighted excess payoff vector is defined, as a result, as $\hat{F}^c(\hat{\mu}) := F^c(\hat{\mu})-\ones F^c(\hat{\mu})^\top\hat{\mu}^c[\Scal^c,\cdot]/m^c$. The players' probability of switching to a policy $v\in \Ucal^c_D$ from any policy, i.e., $\tau^c_v(\hat{F}^c(\hat{\mu}))/\Rrc$, is such that if there exists a strategy with above average payoff, i.e., $\hat{F}(\hat{\mu}) \in \R^{|\Ucal_D|} \setminus \Rnp^{|\Ucal_D|}$, then the expected value of the excess payoff of the transition, i.e., $\tau^c(\hat{F}^c(\hat{\mu}))^\top \hat{F}^c(\hat{\mu}) /\Rrc$, is strictly positive, which is portrayed in Definition~\ref{def:excess}.
	\item  \emph{Pairwise comparison}: When given a revision opportunity, a player of class $c$ using policy $u\in \Ucal^c_D$ compares their payoff, i.e., $F^c_u(\hat{\mu})$, with the payoff of a random policy $v\in \Ucal^c_D$, i.e., $F^c_v(\hat{\mu})$. The switching rates from $u$ to $v$ are positive if and only if the payoff of $v$ strictly exceeds the payoff of $u$, which is portrayed in Definition~\ref{def:pairwise_cmp}.
\end{enumerate} 

For a more detailed discussion on the meaningfulness of these families and of the evolutionary dynamics generated by them refer to \cite[Part~II]{Sandholm2010}. %\notef{Also maybe add a comment saying that different classes may have revision protocols - but same classes used the same (possibly hybrid) protocol}
%\notef{Maybe actually use hybrid protocols like they do in CCW NC Martins. Also the remark there about the protocols being convex cones is nice.}

\begin{mdframed}[style=callout]
	These broad families of revision protocols are characterized by very simple and meaningful qualitative principles. Nevertheless, in reality, players' decisions are complex and multifaceted, and therefore cannot be accurately captured by a single revision protocol from either class. However, if one shows that a game exhibits a certain feature across \emph{all} revision protocols in a class, then one can argue that this feature is induced by the meaningful qualitative principle that characterizes the class. Furthermore, one may design control solutions under the assumption that the specific behavior of the players is unknown but satisfies the meaningful principles defining one of these classes. That endows the control solution with robustness to unavoidable modeling uncertainty.
\end{mdframed}

\subsection{Mean Field Evolutionary Dynamics}

Considering a continuum of players instead of a finite number allows to describe the revision dynamics by the evolution of the joint state-policy distribution of the population. Recall that the joint state-policy distribution at time $t$ is denoted by  $\mu(t) \in X$. We also denote the vector of the mass of a class $c\in [C]$ on each policy as $\mu^c[\Scal^c,\cdot](t) := \col(\mu^c[\Scal^c,u](t),u\in \Ucal^c_D) \in  X^c_{\Ucal_D}$, where the concatenation ordering is consistent with the ordering of $F^c$. Henceforth, the time dependence is oftentimes dropped for conciseness. 

Intuitively, in an infinitesimal interval of time $\dint t$, for a class $c$, the difference in the mass in state $s\in \Scal^c$ evolves according to \eqref{eq:dt_mu_u_S} and the difference in the mass in policy $u\in \Ucal^c_D$: (i)~increases by the proportion of revision clock rings in other policies that switch to policy $u$; and (ii)~decreases by the proportion of revision clock rings in policy $u$ that switch to another policy; i.e., $\forall s\in \Scal^c \; \forall u\in \Ucal^c_D$
\begin{equation*}
	\begin{split}
		\dint \mu^c[s,u] &=   \sum_{s^\prime\in \Scal^c}  \sum_{a^\prime\in \Acal^c(s^\prime)}\Rdc\mu^c[s^\prime,u] \dint t\phi^c(s|s^\prime,a^\prime)u(a^\prime|s^\prime)\\
		&- \Rdc\mu^c[s,u]\dint t \sum_{s^\prime\in \Scal^c} \sum_{a\in \Acal^c(s)}\phi^c(s^\prime|s,a)u(a|s),\\
		 &+ \sum_{u^\prime \in \Ucal_D^c} \Rrc\mu^c[s,u^\prime] \dint t \rho^c_{u^\prime u}(F^c(\mu),\mu^c[\Scal^c,\cdot])/\Rrc \\
		 &- \Rrc \mu[s,u] \dint t\sum_{u^\prime \in \Ucal^c_D} \rho^c_{uu^\prime}(F^c(\mu),\mu^c[\Scal^c,\cdot])/\Rrc.
	 \end{split}
\end{equation*}
When $\dint t \to 0$ this balance equation can be written as 
\begin{equation}\label{eq:ODE_mu_ev}
	\dot{\mu}^c[s,u] =f_{s,u}^{c,d}(\mu) + 	f_{s,u}^{c,r}(\mu),  %\forall c\in [C]\; \forall s\in \Scal^c \;\forall u\in \Ucal^c_D,
\end{equation}
where
\begin{equation}\label{eq:fd_fr}
	\begin{split}
		f_{s,u}^{c,d}(\mu) =&  \Rdc \sum_{s^\prime \in \Scal^c} \sum_{a^\prime \in \Acal^c(s^\prime)}\!\!\!\phi^c(s|s^\prime,a^\prime)u(a^\prime|s^\prime)\mu^c[s^\prime,u] \\-& \Rdc\mu^c[s,u] \\
		f_{s,u}^{c,r}(\mu) =& \sum_{u^\prime \in \Ucal^c_D}  \mu^c[s,u^\prime] \rho^c_{u^\prime u}(F^c(\mu),\mu^c[\Scal^c,\cdot]) \\- &\mu^c[s,u] \sum_{u^\prime \in \Ucal_D^c}  \rho^c_{uu^\prime}(F^c(\mu),\mu^c[\Scal^c,\cdot]).
	\end{split}
\end{equation}
The ODE in \eqref{eq:ODE_mu_ev} is called the \emph{mean dynamic} or \emph{master equation}. Due to the aforementioned regularity assumptions, the mean dynamic is well defined, as formally detailed in the following result.

\begin{lemma}\label{th:ev_ODE_sol}
	Under Assumptions~\ref{ass:cont}-\ref{ass:rev_protocol}, a solution to the master equation, characterized by \eqref{eq:ODE_mu_ev}, with initial condition $\muSU(0)\in X$ exists in $t\in [0,\infty)$, is unique, and is Lipschitz continuous w.r.t.\ $\muSU(0)$. 
\end{lemma}%
\begin{proof}
	First, notice that \eqref{eq:ODE_mu_ev} can be written for all classes $c\in [C]$, states $s\in \Scal^c$ and policies $u\in \Ucal_D^c$ in vector form as an ODE with a vector field $V: X \to T  X$, where $ T  X$ denotes the tangent space of $X$. Second, under Assumption~\ref{ass:markov_unique}, notice that for all $c\in [C]$, all $s\in \Scal^c$ and all $u\in \Ucal^c_D$, $J^c(u,\mu_{\Scal \times \Acal})$, as defined in \eqref{eq:def_J}, can be written as a linear combination of a finite number of single stage reward functions. Therefore, due to Assumption~\ref{ass:cont}, $J^c(u,\mu_{\Scal \times \Acal})$ is Lipschitz continuous w.r.t.\ $\mu$. Hence, for all $c\in [C]$, $F^c(\mu)$, defined in \eqref{eq:def_F}, is Lipschitz continuous w.r.t.\ $\mu$. Furthermore, due to Assumption~\ref{ass:rev_protocol}, $V(\muSU)$ is Lipschitz continuous w.r.t.\ $\mu$. Under these conditions, since $X$ is convex and compact, existence and uniqueness follows from an extension of the Picard-Lindel\"{o}f Theorem to compact convex spaces \cite[Theorem~5.7]{Smirnov2002}\cite[Theorem~4.A.5]{Sandholm2010} and Lipschitz continuity follows from Gr\"onwall's Inequality \cite[Theorem~4.A.3]{Sandholm2010}.
\end{proof}

\begin{theorem}\label{lem:approx_ODE_mu_ev}
	If $\lim_{N\to \infty}\hat{\mu}(0)= \mu(0)$ almost surely, then for all $T<\infty$ $\hat{\mu}(t)$ converges in probability to $\mu(t)$ for all $t\in [0,T]$ as $N\to \infty$.
\end{theorem}
\begin{proof}
	The result follows from Lemma~\ref{th:ev_ODE_sol} and its proof, which allow to directly apply Kurtz's Theorem \cite[Theorems~10.2.1 and~10.2.3]{Sandholm2010}.
\end{proof}

\section{MSNE and Evolutionary Equilibria}\label{sec:rest_point_MSNE}

In this section, we study the relation between a rest point of the evolutionary dynamics \eqref{eq:ODE_mu_ev} and the MSNE solution concept. Due to the way the MSNE is defined, we can build on known results to study that relation. Henceforth, we consider that Assumptions~\ref{ass:cont}-\ref{ass:rev_protocol} hold.

The first result is that every MSNE is a rest point of the evolutionary dynamics for almost all classes of revision protocols defined in Section~\ref{sec:ev_dynamics}. 

\begin{theorem}\label{th:MSNE_is_rest}
	Consider an imitative via comparison, excess payoff, or pairwise comparison revision protocol $\rho^c$ for each class $c\in [C]$. If $\mu \in  X$ is a MSNE, then $\mu$ is a rest point of the evolutionary dynamics \eqref{eq:ODE_mu_ev}.%\notef{This only holds for an hybrid protocol when it does not have any imitative component (except if it is via comparison).}
\end{theorem}
\begin{proof}
	See Appendix~\ref{sec:proof_MSNE_is_rest}.
\end{proof}

Theorem~\ref{th:MSNE_is_rest} does not hold in general if (at least) one class uses imitative revision protocols that are not via comparison. Interestingly, this behavior is different from static games, where a NE is a rest point of the evolutionary dynamics for any imitative revision protocol. Example~\ref{eg:imitative_MSNE_not_rest} below provides insights into this fundamental difference.

\begin{remark}\label{rm:strict_MSNE_is_rest}
	In the particular case whereby $\mu$ is a MSNE for which there is one and only one policy in each class that achieves the maximum payoff, i.e.,  $\argmax_{v\in \Ucal^c_D} F^c_v(\mu)$ has a single element for all $c\in [C]$, then Theorem~\ref{th:MSNE_is_rest} also holds for generic imitative revision protocols. This follows intuitively from the discussion in Example~\ref{eg:imitative_MSNE_not_rest} below and formally from the proof of Theorem~\ref{th:MSNE_is_rest}.
\end{remark}

From Theorem~\ref{th:MSNE_is_rest} it follows that every MSNE is an equilibrium of the evolutionary dynamics under mild conditions. However, for the converse to be true, stronger conditions are required, which are analyzed in what follows.

\begin{theorem}\label{th:rest_is_MSNE}
	Consider an excess payoff or pairwise comparison revision protocol $\rho^c$ for each class $c\in [C]$. If $\muSU \in X$ is a rest point of the evolutionary dynamics \eqref{eq:ODE_mu_ev}, then $\mu$ is a MSNE.%\notef{This only holds for an hybrid protocol if it does not have any imitative component since it requires Lemma~\ref{lem:pc_prop}.}
\end{theorem}
\begin{proof}
	See Appendix~\ref{sec:proof_rest_is_MSNE}.
\end{proof}

Similarly to Theorem~\ref{th:MSNE_is_rest}, Theorem~\ref{th:rest_is_MSNE} does not hold in general  if (at least) one class uses imitative revision protocols. Specifically, for imitative protocols that are \emph{not imitative via comparison} no relation can be established between rest points of the evolutionary dynamics and MSNE. The following example illustrates that a MSNE may not be a rest point and that a rest point may not be a MSNE under these revision protocols.

%In this example, we illustrate that there are imitative revision protocols for which a MSNE is not necessarily a rest point of the evolutionary dynamics and vice versa. 

\begin{example}\label{eg:imitative_MSNE_not_rest}
	Consider a model with a unique class, i.e., $C = 1$, with a state space $\Scal=\{s_1,s_2\}$ and action space $\Acal = \{a_1,a_2\}$, whereby the actions available in state $s_1$ are $\Acal(s_1) = \{a_1\}$ and in state $s_2$ are $\Acal(s_2) = \{a_1,a_2\}$. The state transition matrices upon choosing actions $a_1$ and $a_2$ are given, respectively, by
	\begin{equation*}
		\phi(\cdot|\cdot,a_1) = \begin{bmatrix}
			0.7 & 0.2\\ 0.3 & 0.8
		\end{bmatrix}\quad \text{and} \quad  	\phi(\cdot|\cdot,a_2) = \begin{bmatrix}
		0.5 & 0.7\\ 0.5 & 0.3
	\end{bmatrix}.
	\end{equation*}
	Notice that there are two deterministic policies $\Ucal_D = \{u_1,u_2\}$, which are characterized by $u_1(s_1) = \delta_{a_1}(a)$, $u_1(s_2) = \delta_{a_1}(a)$,  $u_2(s_1) = \delta_{a_1}(a)$, and $u_2(s_2) = \delta_{a_2}(a)$. Consider a revision protocol called \emph{imitation driven by dissatisfaction}, which is an imitative protocol characterized by $r_{uv}(F,\sigma) = (K-F_u)$, where we set $K = 2$. Notice that this revision protocol is imitative, but not imitative via comparison.
	First, consider the state-policy distribution $\mu$ characterized by $\mu[s_1,u_1] = 0.08$, $\mu[s_2,u_1] = 0.12$, $\mu[s_1,u_2] = 0.56$, and $\mu[s_2,u_2] = 0.24$, which is shown in Fig.~\ref{fig:eg_MSNE_not_rest}. Consider that the single-stage reward at $\mu$ is unitary for every state and every action, therefore $F_{u_1}(\mu) = 1$ and $F_{u_2}(\mu) = 1$. Then, $\mu$ is a MSNE according to Definition~\ref{def:MSNE}, since both policies achieve maximum payoff and the state distribution of each policy is stationary. Computing the evolutionary flows according to \eqref{eq:ODE_mu_ev} yields null dynamic flows but nonnull revision flows, which are depicted in Fig.~\ref{fig:eg_MSNE_not_rest}. One concludes that the MSNE $\mu$ is not a rest point.
	Second, consider the state-policy distribution $\mu$ characterized by $\mu[s_1,u_1] = 0.3$, $\mu[s_2,u_1] = 0.3$, $\mu[s_1,u_2] = 0.25$, and $\mu[s_2,u_2] = 0.15$, which is shown in Fig.~\ref{fig:eg_rest_not_MSNE}. Again, consider that the single-stage reward at $\mu$ is unitary for every state and every action, therefore $F_{u_1}(\mu) = 1$ and $F_{u_2}(\mu) = 1$. Then, computing the evolutionary flows according to \eqref{eq:ODE_mu_ev} yields $f^d(\mu)+f^r(\mu) = \zeros$, whose dynamical and revision flows are depicted in Fig.~\ref{fig:eg_rest_not_MSNE}. Despite the fact that both policies achieve maximum payoff, the state distribution of each policy is not stationary, therefore $\mu$ is not a MSNE. One concludes that rest point $\mu$ is not a MSNE.

	\begin{figure}[ht]
		\centering
		\scalebox{0.85}{\begin{tikzpicture}
	% Draw dashed rectangles with thicker lines and margins around the boxes
	\draw[dashed, thick] (-1.75, 2) rectangle (-0.25, -2); % Added margin around the boxes
	\draw[dashed, thick] (2.25, 2) rectangle (3.75, -2); % Added margin around the boxes
	
	% Labels for rectangles (increased y-coordinate for more space)
	\node at (-1, 2.5) {\(u_1\)};
	\node at (3, 2.5) {\(u_2\)};
	
	% Draw square boxes with values with thicker lines
	\node[draw, thick, minimum size=1cm] at (-1, 1.25) {$0.08$};
	\node[draw, thick, minimum size=1cm] at (-1, -1.25) {$0.12$};
	\node[draw, thick, minimum size=1cm] at (3, 1.25) {$0.56$};
	\node[draw, thick, minimum size=1cm] at (3, -1.25) {$0.24$};
	
	% Draw arrows with thicker lines and larger arrowheads
	\draw[<-, thick, >=stealth] (-0.5, 1.25) -- (2.5, 1.25) node[midway, above] {$0.048$};
	\draw[->, thick, >=stealth] (-0.5, -1.25) -- (2.5, -1.25) node[midway, above] {$0.048$};
	
	% Draw vertical dashed arrows with labels and white background (increased height)
	\draw[ thick, <->, >=stealth] (-1.3, 0.75) -- (-1.3, -0.75) node[midway, right, fill=white, minimum height=0.75cm, align=center] {$f^d(\mu) = \zeros$};
	\draw[ thick, <->, >=stealth] (2.7, 0.75) -- (2.7, -0.75) node[midway, right, fill=white, minimum height=0.75cm, align=center] {$f^d(\mu) = \zeros$};
	
	% Additional text
	\node at (-1, -2.5) {\(F_{u_1}(\mu) = 1\)};
	\node at (3, -2.5) {\(F_{u_2}(\mu) = 1\)};
	
	% Move s_1 and s_2 to the left of the respective squares and align vertically
	\node at (-2.3, 1.25) {\(s_1\)};
	\node at (-2.3, -1.25) {\(s_2\)};
\end{tikzpicture}}
		\caption{Example of MSNE $\mu$ that is not a rest point.}\label{fig:eg_MSNE_not_rest}
		\vspace{0.2cm}
		\centering
		\scalebox{0.85}{\begin{tikzpicture}
	% Draw dashed rectangles with thicker lines and margins around the boxes
	\draw[dashed, thick] (-1.75, 2) rectangle (-0.25, -2); % Added margin around the boxes
	\draw[dashed, thick] (2.25, 2) rectangle (3.75, -2); % Added margin around the boxes
	
	% Labels for rectangles (increased y-coordinate for more space)
	\node at (-1, 2.5) {\(u_1\)};
	\node at (3, 2.5) {\(u_2\)};
	
	% Draw square boxes with values with thicker lines
	\node[draw, thick, minimum size=1cm] at (-1, 1.25) {$0.30$};
	\node[draw, thick, minimum size=1cm] at (-1, -1.25) {$0.30$};
	\node[draw, thick, minimum size=1cm] at (3, 1.25) {$0.25$};
	\node[draw, thick, minimum size=1cm] at (3, -1.25) {$0.15$};
	
	% Draw arrows with thicker lines and larger arrowheads
	\draw[<-, thick, >=stealth] (-0.5, 1.25) -- (2.5, 1.25) node[midway, above] {$0.03$};
	\draw[->, thick, >=stealth] (-0.5, -1.25) -- (2.5, -1.25) node[midway, above] {$0.03$};
	
	% Draw vertical dashed arrows with labels and white background (increased height)
	\draw[->, thick, >=stealth] (-1.3, 0.75) -- (-1.3, -0.75) node[midway, right] {$0.03$};
	\draw[<-, thick, >=stealth] (2.7, 0.75) -- (2.7, -0.75) node[midway, right] {$0.03$};
	
	% Additional text
	\node at (-1, -2.5) {\(F_{u_1}(\mu) = 1\)};
	\node at (3, -2.5) {\(F_{u_2}(\mu) = 1\)};
	
	% Move s_1 and s_2 to the left of the respective squares and align vertically
	\node at (-2.3, 1.25) {\(s_1\)};
	\node at (-2.3, -1.25) {\(s_2\)};
\end{tikzpicture}\hspace{0.5cm}}
		\caption{Example of rest point $\mu$ that is not a MSNE.}\label{fig:eg_rest_not_MSNE}
		\vspace{-0.3cm}
	\end{figure}

%		\centering
%		\begin{minipage}{0.95\linewidth}
%			
%		\end{minipage}%
%%		\hspace{0.03\textwidth}
%		\begin{minipage}{0.95\linewidth}
%			\centering
%		
%		\end{minipage}
%	\end{figure}

	In the two cases above, the factor that prevents an equivalence between a MSNE and a rest point is the nonnull revision flow between policies that are payoff maximizing. The class of imitative revision protocols for which revision flows between policies with the same payoff is null is precisely the class of imitative via comparison protocols. Indeed, for imitative via comparison protocols, a MSNE is a rest point (by Theorem~\ref{th:MSNE_is_rest}) and a rest point is a MSNE under additional mild conditions (see Remark~\ref{rm:rest_is_MSNE_im_cmp}). Notice also that, in a case where there is a single policy with maximum payoff, there are no flows between payoff maximizing policies and, as a result, the results that hold for imitative via comparison also hold for general imitative revision protocols. All the code used to generate this example is available in an open-access repository at \weblink{https://github.com/fish-tue/evolutionary-mfg-avg}.
\end{example}
	
It follows from Theorems~\ref{th:MSNE_is_rest} and~\ref{th:rest_is_MSNE}, that, given an excess payoff or pairwise comparison revision protocol, $\mu$ is a MSNE if and only if $\mu$ is an equilibrium point of the evolutionary dynamics \eqref{eq:ODE_mu_ev}. Table~\ref{tab:rev_protocol_properties} summarizes the findings in this section.

\begin{remark}\label{rm:rest_is_MSNE_im_cmp}
	From the intuitive interpretation of imitative dynamics, if a policy $u \in \Ucal^c_D$ of a class $c\in [C]$ does not have any mass in the initial condition, i.e., $\mu^c[\Scal^c,u](0) = 0$, then  $\mu^c[\Scal^c,u](t) = 0$ for all $t\geq 0$. As a result, there can be a rest point of the evolutionary dynamics that does not place mass on a payoff maximizing policy. This observation explains why a rest point of an imitative via comparison revision protocol is not necessarily a MSNE. However, notice that any small perturbation of the revision protocol that places a small mass on such payoff maximizing policy quickly renders the rest point unstable. In Theorem~1 of Part~II of this work, motivated by this observation, we establish that under a very mild Lyapunov stability condition a rest point under an imitative via comparison revision protocol is a MSNE.
\end{remark}

\begin{table}[ht!] \renewcommand{\arraystretch}{1.1}
	\centering
	\caption{{\small Summary of properties of illustrative classes of revision protocols. $^{(\ast)}$In Part~II it is shown that a Lyapunov stable rest point is a MSNE under imitative via comparison revision protocols.}}
	\vspace{-0.1cm}
	\label{tab:rev_protocol_properties} 
	\normalsize
	\begin{tabular}{ll}
		\hline
		Imitative  & MSNE $\centernot\implies$ Rest point \\
		&MSNE \;$\centernot\Longleftarrow$ \;Rest point\\
		Imitative via comparison  & MSNE $\implies$ Rest point\\
		& MSNE\; $\Longleftarrow$ \;Rest point$^{(\ast)}$  \\
		Excess payoff   & MSNE $\iff $ \!Rest point  \\ 
		Pairwise comparison  & MSNE $\iff$ \!Rest point\\ \hline 
	\end{tabular}
\end{table}

%\begin{table}[ht] \renewcommand{\arraystretch}{1.1}
%	\centering
%	\caption{{\small Summary of properties of illustrative classes of revision protocols. $^{(\ast)}$In Part~II it is shown that a Lyapunov stable rest point is a MSNE under imitative via comparison revision protocols.}}
%	\vspace{-0.1cm}
%	\label{tab:rev_protocol_properties}
%	\begin{tabular}{lcc}
%		\hline
%		& MSNE $\implies$ Rest point & Rest point $\implies$  MSNE\\
%		\hline
%		Imitative  & \xmark & \xmark  \\
%		Imitative via comparison  & \cmark & \cmark$^{(\ast)}$  \\
%		Excess payoff   & \cmark & \cmark  \\ 
%		Pairwise comparison  & \cmark &  \cmark\\ \hline 
%	\end{tabular}
%\end{table}
% !TeX spellcheck = en_US
\section{Medium Access Game: Equilibria}\label{sec:MAC_equilibria}

In this section, we illustrate the notions of equilibria resorting to a simple real-life application of a medium access game (MAC) between mobile terminals competing for a common wireless channel. The model used in this section is very similar to the one presented in \cite{WiecekAltmanEtAl2011}. Briefly, each mobile terminal is a player that, from time to time, is required to transmit a message through a common wireless channel. When a player needs to send a message, they choose the level of power at which they want to transmit. The single-stage reward is the signal to interference and noise ratio at the receiver, which depends on the power the message is transmitted at and on the power distribution of the remaining mobile terminals that are using the common channel. Moreover, each player has a battery state that limits the transmission power. Transitions to a lower battery state are more likely the higher the transmission power is.

\subsection{Model}

In this section, we consider a simple version of the MAC which only has one class, three battery states and two transmission power levels. Formally, the mean field model of the MAC, is characterized by: 
\begin{itemize}
	\item \emph{Time}: Each player makes a decision each time a Poisson clock with rate $\Rd$ rings. 
	\item \emph{States}: There are three states $\Scal=\{\sE,\sAE,\sF\}$, corresponding to empty (E), almost empty (AE), and full (F) battery levels.
	\item\emph{Actions}: There are three actions $\Acal = \{\aN,\aL,\aH\}$ corresponding to not transmitting, transmitting at low power, and transmitting at high power. When the battery is empty, no transmission is allowed, i.e., $\Acal(\sE) = \{0\}$; when the battery is almost empty, only low power transmissions are allowed, i.e., $\Acal(\sAE) = \{\aL\}$; and when the battery is full, both low and high power transmissions are allowed, i.e., $\Acal(\sF) = \{\aL,\aH\}$. The transmission powers of actions $\aN$, $\aL$, and $\aH$ are denoted respectively by $P_\aN = 0$, $P_\aL$ and $P_\aH$, which satisfy $0<P_\aL<P_\aH$.
	\item\emph{State transitions}: When a player takes action $\aN$ in state $\sE$ the battery level will be recharged and transition to state $\sF$ with probability $p_\sF$ and to $\sE$ with probability $1-p_\sF$. When a player plays $a \in \{\aL,\aH\}$, the probability of transitioning to the next lower battery state is $\alpha P_a +\gamma$ and of staying in the same energy level is $1-\alpha P_a -\gamma$. Here, $\alpha>0$ and $\gamma>0$ are constants that model the energy consumption due to the transmission of the message and due to other activities, respectively. These constants must satisfy $\alpha P_\aH + \gamma \leq 1$.
	\item \emph{Single-stage reward}: The single-stage reward of a player in state $s$ playing action $a$ when the state-action distribution of the population is $\mu_{\Scal\times \Acal}\in X_{\Scal\times \Acal}$ is the expected signal to interference and noise ratio given by
	\begin{equation*}
		r(s,a,\mu_{\Scal \times \Acal}) \!= \!\frac{P_a}{\sigma^2 + \Rd TC\!\!\!\!\!\sum\limits_{a^\prime \in \{\aL,\aH\}}\!\!\!\!\! P_{a^\prime}\mu_{\Scal \times \Acal}[\Scal,a^\prime]} -\beta P_a,
	\end{equation*}
	where $\sigma, C$, and $\beta$ are constants whose physical interpretation is described in \cite{WiecekAltmanEtAl2011}, and $T$ is the duration of the transmission of a message. Notice that $\Rd T$ is the expected number of clock rings in an interval of $T$ time units, therefore $\Rd T\mu_{\Scal \times \Acal}[\Scal,a]$ is the expected number of messages that are being transmitted with action $a$ at each time instant.
\end{itemize}

Policies in $\Ucal$ are characterized by a scalar $q \in [0,1]$ that represents the probability that a player in state $\sF$ chooses action $\aL$. Specifically, policies in $\Ucal$ are characterized by 
\begin{equation*}
		u_q(s) = \begin{cases}
			\delta_\aN(a), & s = \sE\\
			\delta_\aL(a), & s = \sAE\\
			q\delta_\aL(a) + (1-q)\delta_\aH(a), & s = \sF.
			\end{cases}
\end{equation*}
There exist two deterministic policies, which correspond to the randomized policy $u_q$ when $q = 1$ and $q = 0$, i.e., $\Ucal_D = \{u_1,u_0\}$. Indeed, $u_1$ corresponds to the case where a player deterministically chooses action $\aL$ from state $\sF$ and $u_0$ when a player deterministically chooses action $\aH$ from state $\sF$.

\subsection{BSNE and MSNE}

\begin{figure}[t]
	\centering
	\includegraphics[width=0.85\linewidth]{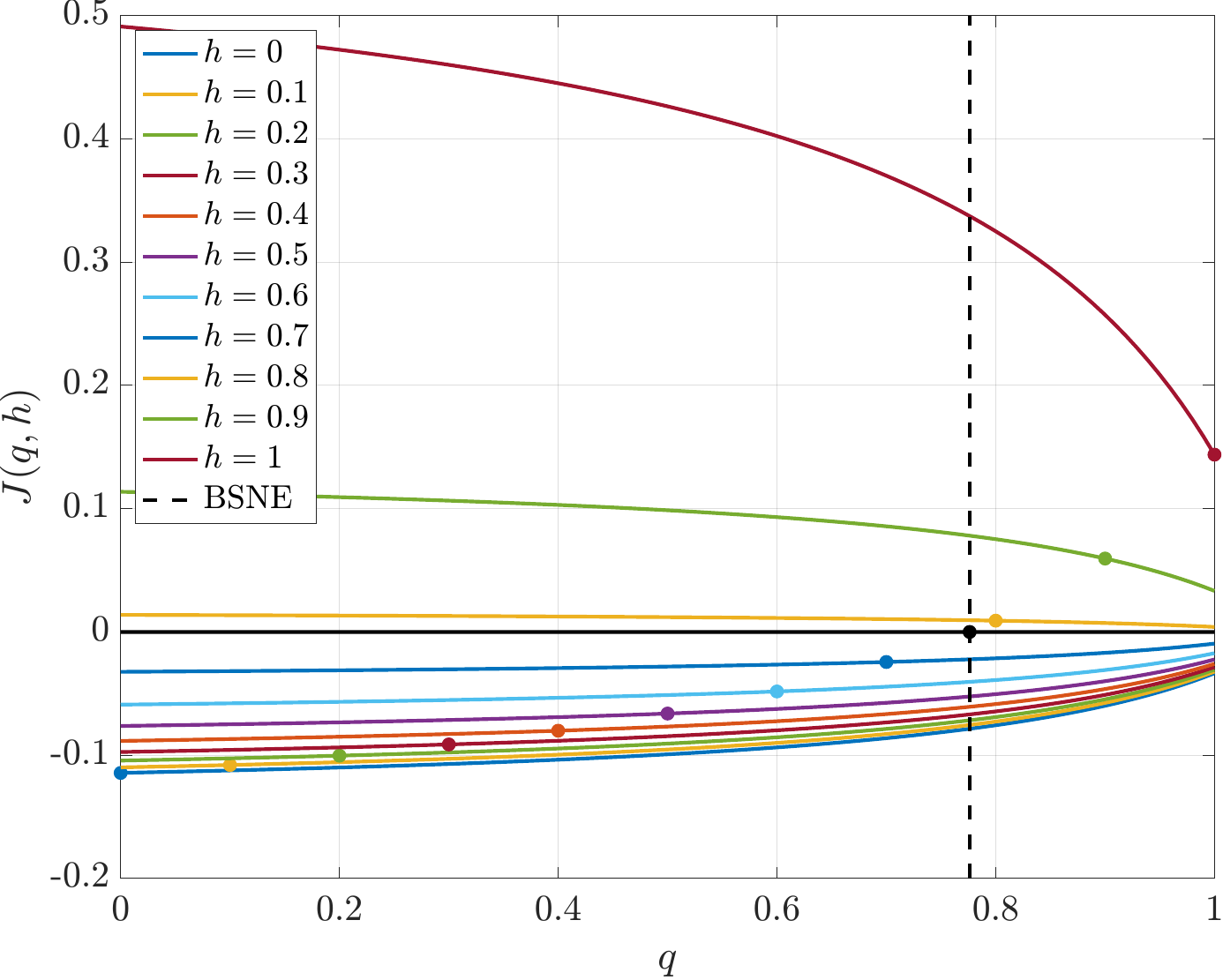}
	\caption{Graphical interpretation of BSNE: For many values of $h$, evolution with $q \in [0,1]$  of the payoff of deviating to policy $u_q$ while the rest of the population uses $u_h$.}
	\vspace{-0.1cm}
	\label{fig:eg_mac3_bsne}
	% BSNE = 0.78
\end{figure}

First, recall from Section~\ref{sec:equilibria} that a BSNE is characterized by a randomized policy in $\Ucal$ and the corresponding stationary state distribution such that no player can unilaterally deviate to increase their payoff. Consider the whole population is playing $u_h \in \Ucal$ and that a player unilaterally deviates to $u_q \in \Ucal$. The long-time average payoff of the player that deviates is given by
\begin{equation*}\small
	\begin{split}
	&J(q,h) = \left((\eta_q(\sAE)+q\eta_q(\sF))P_\aL + (1-q)\eta_q(\sF)P_\aH \right)\\
	&\left(\frac{1}{\sigma^2 \!+ \!\Rd TC ((\eta_q(\sAE)+q\eta_q(\sF))P_\aL + (1\!-\!q)\eta_q(\sF)P_\aH )} -\beta\right)\!,
	\end{split}
\end{equation*}   
where $\eta_q$ and $\eta_h$ are the unique stationary state distributions of using policies $u_q$ and $u_h$, respectively, which exist since the Markov jump chain associated with the state transitions of any policy in $\Ucal$ is irreducible. Fig.~\ref{fig:eg_mac3_bsne} depicts the evolution of $J(q,h)$ with $q$ for many values of $h$ for randomly chosen parameters. Notice along the lines with $h\leq 0.7$ a player can unilaterally deviate from $q = h$ to $q>h$ to increase their payoff. Similarly, along the lines $h\geq 0.8$ a player can unilaterally deviate from $q = h$ to $q<h$ to increase their payoff. At $h = h^\star \approx 0.78$, depicted in black in Fig.~\ref{fig:eg_mac3_bsne}, no player can deviate from $q = h^\star$ to $q\neq h^\star$ to increase their payoff. Therefore, the pair $(u_{h^\star},\eta_{h^\star})$ is a BSNE.

\begin{figure}[t]
	\centering
	\includegraphics[width=0.82\linewidth]{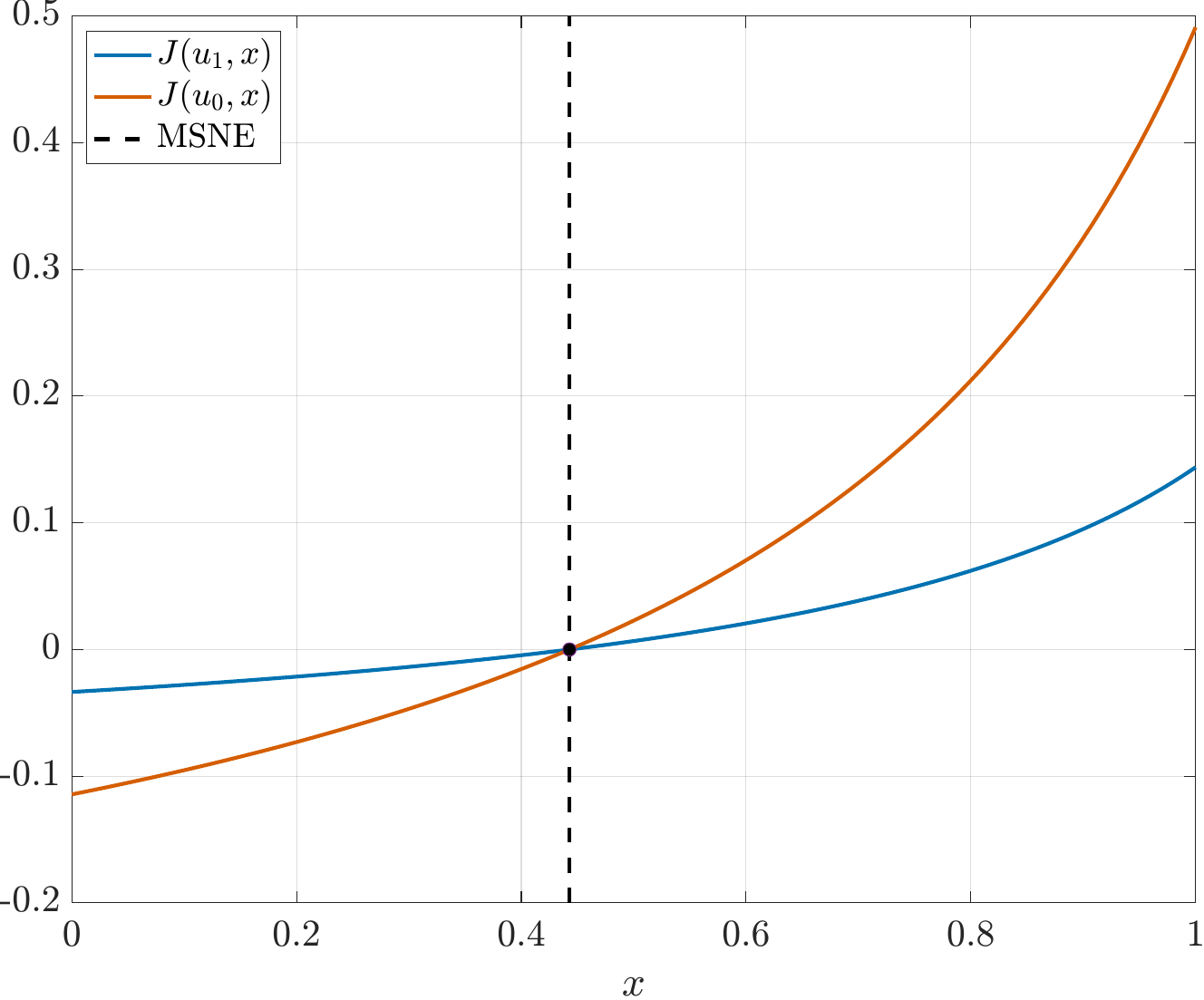}
	\caption{Graphical interpretation of MSNE: Evolution with $x\in [0,1]$ of the payoff of playing policies $u_1$ and $u_0$ while the proportion of the remainder of the population playing $u_1$ is $x$ and playing $u_0$ is $1-x$.}
	\label{fig:eg_mac3_msne}
	\vspace{-0.1cm}
	% MSNE = 0.44
\end{figure}

Second, recall that a MSNE is characterized by a joint state-policy distribution, whereby the state distribution of each fixed policy is stationary, each player uses a deterministic policy, and no player can unilaterally deviate to another deterministic policy to obtain a better payoff. Consider that the proportion of the population playing $u_1$ is denoted by $x$. The long-time average payoff of a player using $u_1 \in \Ucal_D$ is given by
\begin{equation*}
	\begin{split}
	J(u_1,x) = \left(\eta_1(\sAE)+\eta_1(\sF) \right)P_\aL \left(\frac{1}{ \sigma^2 \!+ \!\Rd TC\bar{P}(x) } \!-\! \beta \right),
	\end{split}
\end{equation*} 
where $\bar{P}(x) = (x(\eta_1(\sAE)+\eta_1(\sF))+ (1-x)\eta_0(\sAE))P_\aL + (1-x)\eta_0(\sF)P_\aH$,
and of a player using $u_0 \in \Ucal_D$ is given by
\begin{equation*}
	\begin{split}
	J(u_0,x) =  \left(\eta_0(\sAE)P_\aL \!+ \! \eta_0(\sF)P_\aH \right) \left(\frac{1}{ \sigma^2 \!+ \!\Rd TC\bar{P}(x) } \!-\! \beta \right)\!.
	\end{split}
\end{equation*}%
Fig.~\ref{fig:eg_mac3_msne} depicts the evolution of $J(u_1,x)$ and $J(u_0,x) $ with $x$. Notice that for $x<x^\star \approx 0.44$, a player using policy $u_0$ can unilaterally change to $u_1$ to increase their payoff. Similarly, for $x>x^\star$ a player using policy $u_0$ can unilaterally change to $u_1$ to increase their payoff. At $x = x^\star$, depicted in a vertical dashed line  in Fig.~\ref{fig:eg_mac3_bsne}, no player can deviate from either policy to increase their payoff. Therefore, $\mu^\star \in X$ characterized by $\mu^\star[\sE,u_1] = x^\star\eta_1(\sE)$, $\mu^\star[\sAE,u_1] = x^\star\eta_1(\sAE)$, $\mu^\star[\sF,u_1] = x^\star\eta_1(\sF)$, $\mu^\star[\sE,u_0] = (1-x^\star)\eta_0(\sE)$, $\mu^\star[\sAE,u_0] = (1-x^\star)\eta_0(\sAE)$, $\mu^\star[\sF,u_0] = (1-x^\star)\eta_0(\sF)$ is a MSNE.

%\vspace{-0.2cm}

%\subsection{Discussion}
%It is interesting to discuss three points that are illustrated by the MAC example. 

First, notice that the proportion of players choosing action $\aL$ from state $\sF$ at the BSNE is approximately $0.78$, while at the MSNE it is approximately $0.44$. %This illustrates the difference between the payoff when actions of each player are randomized for each state (in the BSNE) and when the actions of each player are deterministic for each state (in the MSNE).
Second, the way the population may reach a MSNE has an evolutionary interpretation. Consider that the policy distribution of the population is in a certain state $x<x^\star$. Intuitively (and ignoring the effect of state transitions), if a proportion of the population is given the opportunity to revise their policy, the revising players that are using policy $u_0$ will realize that they can increase their payoff by switching to strategy $u_1$, as visible in Fig.~\ref{fig:eg_mac3_msne}. So they will switch with some probability, thereby increasing $x$. After many revision opportunities the state of the population will increase until it approaches the MSNE at $x^\star$. The analysis is similar if the initial policy distribution is  $x<x^\star$. However, the way the population approaches the BSNE does not have such evolutionary interpretation. Indeed, to switch from using a policy $u_q \in \Ucal$ to another $u_{q^\prime} \in \Ucal$ the \emph{whole population has to agree to change the way they randomize their actions}, which is not physically meaningful through an evolutionary lens.
Third, despite the fact that we know that $\mu^\star$ is a MSNE, one cannot conclude on the stability of the MSNE with the analysis tools presented thus far. Part~II \cite{PedrosoAgazziEtAl2025MFGAvgII} of this work addresses this aspect.
All the code used the MAC example is available in an open-access repository at \weblink{https://github.com/fish-tue/evolutionary-mfg-avg}.

%\vspace{-0.2cm}

% !TeX spellcheck = en_US
\section{Conclusion}\label{sec:conclusion}

In Part~I of this work, for the first time in the literature, we propose an evolutionary model for the class of continuous-time finite-state stochastic dynamic games of many players. First, we conclude that the finite population game can be approximated with strong guarantees by a mean field approximation, whose simplicity allows for a deeper qualitative analysis. Second, we conclude that the state-of-the-art solution concepts for this class of games do not have an evolutionary interpretation. We propose a new solution concept, which we call mixed stationary Nash Equilibrium (MSNE), that does. Third, the main results of this part indicate that there is an equivalence relation between the proposed MSNE solution concept and the equilibrium points of the mean field evolutionary dynamics. Crucially, the equivalence holds under whole classes of meaningful revision protocols. Fourth, it is important to stress that the mean field approximation of the dynamic game is \emph{not generally suitable for numerical computation of equilibria or trajectories of the game}. The reason is that the cardinality of the set of deterministic policies grows exponentially w.r.t. the number of states. The usefulness of the mean field approximation is that it unlocks a qualitative analysis of the behavior of the players through an evolutionary lens and paves the way for tractable prescription of equilibria.

All in all, if one designs a dynamic game such that a desired population state is a MSNE, the analysis of Part~I allows to establish that such population state is a rest point of meaningful evolutionary dynamics. However, to guarantee the long-term viability of MSNE, i.e., that MSNE can robustly emerge and persist against strategic deviations, requires a stability analysis of the evolutionary model. Such an endeavor is the focus of Part~II \cite{PedrosoAgazziEtAl2025MFGAvgII} of this work.

\appendix
% !TeX spellcheck = en_US

\section{Proofs of Sections~\ref{sec:model} and \ref{sec:equilibria}}

\subsection{Proof of Lemma~\ref{lem:approx_ODE_mu_u_S}}\label{sec:proof_lem_approx_ODE_mu_u_S}
For each $c\in [C]$ and each $u\in \Ucal_D^c$, \eqref{eq:ODE_mu_u_S} can be written for all states in vector form as an ODE whose vector field is Lipschitz continuous and lies on $\{\nu\in \R^{p^c}: \ones^\top\nu = 0\}$.  Existence and uniqueness follow from an extension of the Picard-Lindel\"{o}f Theorem to compact convex spaces \cite[Theorem~5.7]{Smirnov2002}\cite[Theorem~4.A.5]{Sandholm2010} and Lipschitz continuity follows from Gr\"onwall's Inequality \cite[Theorem~4.A.3]{Sandholm2010}. For each fixed $u\in \Ucal_D^c$, we are in the conditions of  Kurtz's Theorem \cite[Theorem~2.1 in Chap.~11]{EthierKurtz1986}, which allows to conclude that $\lim_{N\to \infty}\hat{\mu}^c(t)= \mu^c(t)$ almost surely for all $t\in [0,\infty)$. The map between state-policy distributions and state-action distributions is continuous, so it follows from the continuous mapping theorem~\cite{MannWald1943} that $\lim_{N\to \infty}\hat{\mu}^c_{\Scal \times \Acal}(t)= \mu^c_{\Scal \times \Acal}(t)$ almost surely for all $t\in [0,\infty)$.

\vspace{-0.3cm}

\subsection{Proof of Lemma~\ref{lem:markov_unique}}\label{sec:proof_lem_markov_unique}
The proof consists of two parts. First, we show that, under Assumption~\ref{ass:markov_unique}, for any $c\in [C]$ and any $u\in \Ucal^c$, the continuous-time Markov chain generated by $Q^{c,u}$ has one and only one recurrent communicating class.  For that, notice that there is a deterministic policy $u^\prime \in \Ucal^c_D$ whose non-null probability state transitions are a subset of the non-null state transitions of $u$. By Assumption~\ref{ass:markov_unique}, the state transition Markov chain of $u^\prime$ has one and only one communicating class. Now, notice that modifying the Markov chain associated with $u^\prime$ by introducing a state transition with non-null probability does not increase the number of recurrent communicating classes. That is because a transient state can only become recurrent if it can reach and be reached by a recurrent state of the original chain of $u^\prime$, which just extends a communicating class. By an induction argument, one can sequentially add non-null state transitions to the chain associated with $u^\prime$ to obtain the chain associated with $u$, hence one concludes that $Q^{c,u}$ has one and only one recurrent communicating class. Second, consider a player $i\in \Ccal_c$ whose initial state distribution is any $\eta \in \Pcal(\Scal^c)$, i.e., $s^{i}(0) \sim \eta$. Consider any state $s\in\Scal^c$ and consider two cases: (i)~$s$ is transient; and (ii)~$s$ is recurrent. In case (i), it follows from the definition of a transient state that the total time spent in $s$ is finite, therefore $\mathbb{P}( \lim_{t\to \infty}\frac{1}{t}\int_{0}^t \delta_{s}(s^{i}(\tau)) \dint \tau = 0) \geq \mathbb{P}(\lim_{t\to \infty}\frac{1}{t}\int_{0}^\infty \delta_{s}(s^{i}(\tau)) \dint \tau = 0 )  = 1$.
%\begin{equation*}
%	\mathbb{P}\left( \lim_{t\to \infty}\frac{1}{t}\int_{0}^t \delta_{s}(s^{i,N}(\tau)) \dint \tau = 0\right) \geq \mathbb{P}\left(\lim_{t\to \infty}\frac{1}{t}\int_{0}^\infty \delta_{s}(s^{i,N}(\tau)) \dint \tau = 0 \right)  = 1.
%\end{equation*}
One concludes that the long-time mass on transient states is null almost surely and, as a result, an invariant measure places null mass on transient states. Therefore, the proof of the result reduces to the analysis of case~(ii). In case (ii), since there is one and only one recurrence class, $s^{i}(t)$ hits $s$ with probability one and, since the state space is finite, then $s$ is positive recurrent, i.e., the expected return time is finite. Therefore, we are in the conditions of \cite[Theorem~3.5.2]{Norris1997} and \cite[Theorem~3.8.1]{Norris1997}, which immediately prove the result.

\vspace{-0.2cm}

\subsection{Proof of Theorem~\ref{th:approx_MSNE}}\label{sec:proof_th_approx_MSNE}

%\textcolor{red}{[This proof will change after discussion with Andrea]}

The following proposition is the key to proving the result.
\begin{proposition}\label{propo:aux_approx_MSNE}
	Under the conditions of Theorem~\ref{th:approx_MSNE}, for any player $i\in [N]$,	$\lim_{N\to \infty} J^{i,N}(u^1,u^2,\ldots,u^i,\ldots,u^N) = F^{c^i}_{u^i}(\mu)$.
\end{proposition}
\begin{proof}
	First, we show that for all $i \in [N]$ the convergence of $\lim_{T\to \infty}\frac{1}{T}\EV [\sum_{k=1}^T r(s^i(t^i_k),a^i(t^i_k),\hat{\mu}_{\Scal \times \Acal}(t^i_k))] $ is uniform in $N$. By Assumption~\ref{ass:markov_unique} and the finiteness of $\Scal$, for every policy $u\in \Ucal_D^c$ of any class $c\in [C]$, there exist $k_u \in \mathbb N$,  $\epsilon_u > 0$ and a probability measure $q$ on $\Scal$
	such that $(\phi^{c,u})^{k_u}_{s,\cdot}\ge \varepsilon_u\, q(\cdot)$ for all $s\in \Scal$, where the matrix $(\phi^{c,u})^{k}$ denotes the $k$-fold product of $\phi^{c,u}$.
	This Doeblin minorization condition, combined with Lemma~\ref{lem:markov_unique} guaranteeing the existence and uniqueness of an invariant measure $\eta^{c,u} \in \mathcal P(\Scal)$, implies geometric ergodicity, i.e.,$\|(\phi^{c,u})^{n k_u}\mu_0 -\eta^{c,u}\|_{\mathrm{TV}}\le (1-\epsilon_u)^n$ for all $\mu_0\in \mathcal P(\Scal)$ and all $n \in \N$, where $||\cdot||_{\mathrm{TV}}$ denotes the total variation norm for probability measures.
	One concludes that, for all $j\in [N]$, $s^j(t)$ converges in distribution exponentially fast as $t \to \infty$ and uniformly in $N$. Since $\hat{\mu}_{\Scal \times \Acal}$ is characterized by $\hat{\mu}^c_{\Scal \times \Acal}[s,a](t) = \frac{1}{N}\sum_{j\in \Ccal_c} \delta_{s^{j}(t)}(s) \delta_{u^{j}(s^j(t))}(a)$, it follows that $r(s^i(t),u^i(s^i(t),\hat{\mu}_{\Scal \times \Acal}(t))$ can be written as a function of the r.v.s $s^j(t)$ with $j\in [N]$. Since the single-stage reward is continuous and bounded by Assumption~\ref{ass:cont}, it follows from the Portmanteau theorem \cite[Therorem~10.1.1]{Rosenthal2006} that $\EV\left[r\big(s^i(t),u^i(s^i(t)),\hat\mu_{\Scal \times \Acal}(t)\big)\right]$ converges as $t\to \infty$ uniformly in $N$, which establishes the statement.
	
%		\[
%	\|(\phi^{c,u})^{n k_u}\mu_0 -\eta^{c,u}\|_{\mathrm{TV}}\le (1-\epsilon_u)^n,
%	\]
	
	Second, by Lemma~\ref{lem:approx_ODE_mu_u_S}, for any $t\geq 0$, $\hat{\mu}_{\Scal \times \Acal}(t)$ converges to the mean-field distribution $\mu_{\Scal \times \Acal}(t)$ with probability one. Therefore, applying the Dominated Convergence Theorem \cite[Theorem~9.1.2]{Rosenthal2006} yields
	\begin{equation*}
		\begin{split}
		\lim_{N\to \infty}\frac{1}{T}&\EV\left[\sum\nolimits_{k=1}^T r(s^i(t^i_k),a^i(t^i_k),\hat{\mu}_{\Scal \times \Acal}(t^i_k))\right] \\= \frac{1}{T}&\EV\left[\sum\nolimits_{k=1}^T r(s^i(t^i_k),a^i(t^i_k),\mu_{\Scal \times \Acal}(t^i_k))\right].
	\end{split}
	\end{equation*}
	
	Finally, since  the convergence of $\lim_{T\to \infty} \frac{1}{T} \EV[\sum_{k=1}^T r(s^i(t^i_k),a^i(t^i_k),\hat{\mu}_{\Scal \times \Acal}(t^i_k))]$ is uniform in $N$ and $\lim_{N\to \infty} \frac{1}{T} \EV[\sum_{k=1}^T r(s^i(t^i_k),a^i(t^i_k),\hat{\mu}_{\Scal \times \Acal}(t^i_k))]$ exists, using the Moore-Osgood theorem \cite[Chap.~4, Sec.~11, Theorem~2]{Zakon2004}, one can interchange the limits in $N$ and $T$, i.e., 
	\begin{equation}\label{eq:aux_interchange_limits}
		\begin{split}
			&\lim_{N\to \infty} J^{i,N}( u^1, u^2, \ldots,u^i, \ldots, u^N) \\= & \! \lim_{N\to \infty } \lim_{T\to \infty}\!\frac{1}{T}\EV\!\left[\sum\nolimits_{k=1}^T \!r(s^i(t^i_k),a^i(t^i_k),\hat{\mu}_{\Scal \times \Acal}(t^i_k))\right]\\
			=& \!\lim_{T\to \infty } \lim_{N\to \infty}\!\frac{1}{T}\EV\!\left[\sum\nolimits_{k=1}^T \!r(s^i(t^i_k),a^i(t^i_k),\hat{\mu}_{\Scal \times \Acal}(t^i_k))\right]\\
			=&  \lim_{T\to \infty }  \frac{1}{T}\EV\left[\sum\nolimits_{k=1}^T r(s^i(t^i_k),a^i(t^i_k),\mu_{\Scal \times \Acal}(t^i_k))\right].
		\end{split}
	\end{equation}
	From Lemma~\ref{lem:markov_unique}, the state distribution of a player $i$ converges with probability one to $\eta^{u^i}$ as $k\to \infty$. Moreover, the deterministic mean field state-action distribution $\mu^c_{\Scal \times \Acal}(t)$ converges to $\mu_{\Scal \times \Acal}^{c,\infty} \in \Pcal(\Scal\times \Ucal_D)$ for all $c\in [C]$ as $t\to \infty$ by \eqref{eq:MSNE_s} and \eqref{eq:cond_N_dist_pol}, where $\mu_{\Scal \times \Acal}^{c,\infty}$ is characterized by 
	\begin{equation*}
		\mu_{\Scal\times \Acal}^{c,\infty}[s,a] =\!\!\!\sum_{u\in \Ucal^c_D} \mu^c[s,u]\eta^u(s)u(a|s) = \!\!\! \sum_{u\in \Ucal_D^c} \mu^c[\Scal^c,u]u(a|s)
	\end{equation*}
	for all $s\in \Scal^c$, all $a\in \Acal^c$, and all $c\in [C]$. Hence, applying the Dominated convergence Theorem \cite[Theorem~9.1.2]{Rosenthal2006} to \eqref{eq:aux_interchange_limits} yields 
	\begin{equation*}
		\begin{split}
			&\lim_{N\to \infty} J^{i,N}( u^1, u^2, \ldots,u^i, \ldots, u^N)  \\
			=&  \sum_{s\in\Scal^c} \sum_{a\in \Acal^c(s)} \eta^{u^i}(s)u^i(a|s)r(s,a,\mu^\infty_{\Scal \times \Acal}) = F^{c^i}_{u^i}(\mu),
		\end{split}
	\end{equation*}
	where $F^{c^i}_{u^i}(\mu)$ is defined as in \eqref{eq:def_F}.
\end{proof}
For all $c\in [C]$ and all $i\in [N]$, by condition \eqref{eq:cond_N_dist_pol}, it follows that $\mu^c[\Scal^c,u^i]>0$. Therefore, since $\mu$ is a MSNE by hypothesis, one concludes from the definition of a MSNE in Definition~\ref{def:MSNE} that $F^{c^i}_{u^i}(\mu) = \max_{v\in \Ucal^{c^i}_D}F^{c^i}_v(\mu)$. Therefore, from Proposition~\ref{propo:aux_approx_MSNE},
\begin{equation}\label{eq:aux_approx_MSNE_conv1}
	\lim_{N\to \infty} J^{i,N}( u^1, u^2, \ldots,u^i, \ldots, u^N)  = \max_{v\in \Ucal^{c^i}_D}F^{c^i}_v(\mu).
\end{equation}
Moreover, using the same arguments, one concludes that when player $i$ uses any $v^i\in \Ucal_D^{c^i}$
\begin{equation}\label{eq:aux_approx_MSNE_conv2}
	\!\!\lim_{N\to \infty} \!J^{i,N}\!(u^1, \ldots,v^i, \ldots, u^N)  \! \leq \! \max_{v\in \Ucal_D^{c^i}} \!F^{c^i}_v(\mu),  \forall v\!\in\! \Ucal_D.
\end{equation}
Hence, by \eqref{eq:aux_approx_MSNE_conv1} and \eqref{eq:aux_approx_MSNE_conv2}, $ \lim_{N\to \infty} \!J^{i,N}\!( u^1, \ldots,v^i, \ldots, u^N)  \leq 	\lim_{N\to \infty} \!J^{i,N}\!( u^1, \ldots,u^i, \ldots, u^N),$
for all $v\in \Ucal^{c^i}_D$. Therefore, by the definition of limit, for any $\epsilon>0$ there is $N_\epsilon \in \N$ such that for all $N>N_\epsilon$, $J^{i,N}( u^1, u^2, \ldots,u^i, \ldots, u^N) > J^{i,N}( u^1, u^2, \ldots,v^i, \ldots, u^N) -\epsilon$ for all $v\in \Ucal^{c^i}_D$. One concludes from Definition~\ref{def:MSNE_finite} that $\{u^i\}_{i\in [N]}$ is a weak $\epsilon$-MSNE in the average payoff finite-population game.

\section{Proofs of Section~\ref{sec:rest_point_MSNE}}

%\subsection{Proof of Theorem~\ref{th:ev_ODE_sol}}\label{sec:proof_ev_ODE_sol}
%First, notice that \eqref{eq:ODE_mu_ev} can be written for all states $s\in \Scal$ and policies $u\in \Ucal_D$ in vector form as an ODE with a vector field $V: \Pcal(\Scal\times \Ucal_D) \to T  \Pcal(\Scal\times \Ucal_D)$, where $ T  \Pcal(\Scal\times \Ucal_D)$ denotes the tangent space of $\Pcal(\Scal\times \Ucal_D)$. Second, notice that for all $s\in \Scal$ and all $u\in \Ucal_D$, $J(u,s,\mu_{\Scal \times \Acal})$, as defined in \eqref{eq:def_J}, can be written as a linear combination of a finite number of single state reward functions. Therefore, due to Assumption~\ref{ass:cont}, $J(u,s,\mu_{\Scal \times \Acal})$ is Lipschitz continuous w.r.t.\ $\muSU$. Hence, for any $s\in \Scal$, $F^s(\muSU)$, defined in \eqref{eq:def_Fs}, is Lipschitz continuous w.r.t.\ $\muSU$. Furthermore, due to Assumption~\ref{ass:rev_protocol}, $V(\muSU)$ is Lipschitz continuous w.r.t.\ $\muSU$. Under these conditions, since $\Pcal(\Scal\times \Ucal_D)$ is convex and compact, existence and uniqueness follows from an extension of the Picard-Lindel\"{o}f Theorem to compact convex spaces \cite[Theorem~5.7]{Smirnov2002}\cite[Theorem~4.A.5]{Sandholm2010} and Lipschitz continuity follows from Gr\"onwall's Inequality \cite[Theorem~4.A.3]{Sandholm2010}.

\vspace{-0.3cm}

\subsection{Proof of Theorem~\ref{th:MSNE_is_rest}}\label{sec:proof_MSNE_is_rest}

Throughout the proof define the set of optimal policies of class $c\in [C]$ at $\mu$ by $\Ucal_D^{c\star}(\mu) :=  \argmax_{v\in \Ucal^c_D}F^c_v(\mu)$. The following lemmas establish properties that will be instrumental in the proofs of a few results. The first establishes known results in the context of this problem.

\begin{lemma}\label{lem:pc_ns}
	Let $\rho^c$ be a revision protocol and $\mu \in X$. Consider the following statements:
	\begin{enumerate}[(i)]
		\item $\mu^c[\Scal^c,u]> 0 \implies u\in \Ucal_D^{c\star}(\mu)$ for all $u\in \Ucal^c_D$;
		\item $\sum_{s\in\Scal^c} f^{c,r}_{s,u}(\mu) = 0$ for all $u\in \Ucal^c_D$.
	\end{enumerate}	
	If $\rho^c$ is an imitative, excess payoff, or pairwise comparison revision protocol, then (i)$\implies$(ii). If $\rho^c$ is an excess payoff, or pairwise comparison revision protocol, then (ii)$\implies$(i).%\notef{Add extension to hybrid protocol? First, (i)$\implies$(ii) is PC so satisfied by all hybrid. Second, (ii)$\implies$(i) is NS so satisfied by all hybrid that are not fully imitative.}
\end{lemma}
\begin{proof}
	The implication (i)$\implies$(ii) follows from a property called positive correlation that is satisfied by imitative \cite[Theorems~5.4.9]{Sandholm2010}, excess payoff  \cite[Theorems~5.5.2]{Sandholm2010}, and pairwise comparison  \cite[Theorems~5.6.2]{Sandholm2010} revision protocols. It follows from \cite[Proposition~5.2.1]{Sandholm2010} that if $\mu$ satisfies (i), then $\sum_{u^\prime \in \Ucal^c_D}  \mu^c[\Scal^c,u^\prime] \rho^c_{u^\prime u}(F^c(\mu),\mu^c[\Scal^c,\cdot]) - \mu^c[\Scal^c,u] \sum_{u^\prime \in \Ucal^c_D}  \rho^c_{uu^\prime}(F^c(\mu),\mu^c[\Scal^c,\cdot]) = 0$ for all $u\in \Ucal^c_D$. From \eqref{eq:fd_fr}, it follows that $\sum_{s\in\Scal^c} f^{c,r}_{s,u}(\mu) = 0$ for all $u\in \Ucal^c_D$. The implication (ii)$\implies$(i) follows from a property called Nash stationarity that is satisfied by excess payoff  \cite[Theorems~5.5.2]{Sandholm2010} and pairwise comparison  \cite[Theorems~5.6.2]{Sandholm2010} revision protocols. As a result, it follows that if statement~(ii) holds, i.e.,  $\sum_{u^\prime \in \Ucal^c_D}  \mu^c[\Scal^c,u^\prime] \rho^c_{u^\prime u}(F^c(\mu),\mu^c[\Scal^c,\cdot]) - \mu^c[\Scal^c,u] \sum_{u^\prime \in \Ucal^c_D}  \rho^c_{uu^\prime}(F^c(\mu),\mu^c[\Scal^c,\cdot]) = 0$ for all $u\in \Ucal^c_D$ for all $u\in \Ucal^c_D$, then~(ii) holds.	
\end{proof}

\begin{lemma}\label{lem:pc_prop}
	Consider an imitative via comparison, excess payoff, or pairwise comparison revision protocol $\rho^c$. If $\mu \in X$ satisfies $\mu^c[\Scal^c,u]> 0 \implies u\in \Ucal_D^{c\star}(\mu)$ for all $u\in \Ucal^c_D$, then:%\notef{This only holds when the hybrid does not have any imitative component.}
	\begin{enumerate}[(i)]
%		\item $\sum_{s\in\Scal} f^r_{s,u}(\mu) = 0$ for all $u\in \Ucal_D$;
		\item $\rho^c_{u,v}(F^c(\mu),\mu^c[\Scal^c,\cdot]) = 0\;$ for all $u,v\in\Ucal_D^{c\star}(\mu)$;
		\item $f^{c,r}_{s,u}(\mu) = 0\;$ for all $s\in \Scal^c$ and $u\in \Ucal^c_D$.
	\end{enumerate}
\end{lemma}
\begin{proof}
	To prove statement~(i), notice that if $u,v\in \Ucal_D^{c\star}(\mu)$, then $F^c_u(\mu) =F^c_v(\mu)$. By the definition of imitative via comparison and pairwise comparison revision protocols in Definitions~\ref{def:imitative_cmp} and~\ref{def:pairwise_cmp}, respectively, it follows immediately that $\rho^c_{u,v}(F^c(\mu),\mu^c[\Scal^c,\cdot]) = 0$. For excess payoff revision protocols, albeit not clear from the definition, it follows from continuity of $\rho^c$ that $\rho^c_{u,v}(F^c(\mu),\mu[\Scal^c,\cdot]) = 0$~\cite[Exercise~5.5.7~(ii)]{Sandholm2010}. To prove statement~(ii), we treat two cases separately: (a)~$u\notin \Ucal_D^{c\star}(\mu)$; and (b)~$u\in \Ucal_D^{c\star}(\mu)$. First, notice that, in case~(a), $\mu^c[\Scal^c,u] = 0$ and therefore it follows from the definition of $f^{c,r}_{s,u}(\mu)$ in \eqref{eq:fd_fr} that
	\begin{equation}\label{eq:proof_aux_fr}
		f^{c,r}_{s,u}(\mu) = \sum_{v \in \Ucal^c_D}  \mu^c[s,v] \rho^c_{v u}(F^c(\mu),\mu^c[\Scal^c,\cdot]),
	\end{equation}
	so $f^{c,r}_{s,u}(\mu) \geq 0$ for all $s\in \Scal^c$. Furthermore, it follows from Lemma~\ref{lem:pc_ns} that if $\mu$ is in the conditions of this lemma, then $\sum_{s\in\Scal^c} f^{c,r}_{s,u}(\mu) = 0$ for all $u\in \Ucal^c_D$. Therefore, since $\sum_{s\in\Scal^c} f^{c,r}_{s,u}(\mu) = 0$ and $f^{c,r}_{s,u}(\mu) \geq 0$, it follows that $f^{c,r}_{s,u}(\mu) = 0$ for all $u\notin  \Ucal_D^{c\star}(\mu)$ and all $s\in \Scal^c$.
	Second, we address case~(b). From \eqref{eq:proof_aux_fr} and $f^{c,r}_{s,u^\prime}(\mu) = 0$ for $u^\prime \notin \Ucal_D^{c\star}(\mu)$ it follows that $\forall s\in \Scal^c \; \forall u\in \Ucal_D^c \;\forall u^\prime \notin \Ucal_D^{c\star}(\mu)$
	\begin{equation}\label{eq:proof_aux_fr_null}
		\mu^c[s,u] = 0 \;\lor \; \rho^c_{u u^\prime}(F^c(\mu),\mu^c[\Scal^c,\cdot]) = 0. 
	\end{equation}
	Expanding the expression for $f^{c,r}_{s,u}(\mu)$ in \eqref{eq:fd_fr} with $u\in \Ucal_D^{c\star}(\mu)$ yields
	\begin{equation*}
		\begin{split}
			f_{s,u}^{c,r}(\mu) =& \sum_{u^\prime \notin  \Ucal_D^{c\star}(\mu)}  \mu^c[s,u^\prime] \rho^c_{u^\prime u}(F(\muSU),\mu^c[\Scal^c,\cdot])\\
			- & \mu^c[s,u] \sum_{u^\prime \notin  \Ucal_D^{c\star}(\mu) }  \rho^c_{uu^\prime}(F^c(\mu),\mu^c[\Scal^c,\cdot])\\
			+ & \sum_{u^\prime \in  \Ucal_D^{c\star}(\mu)}  \mu^c[s,u^\prime] \rho^c_{u^\prime u}(F^c(\mu),\mu^c[\Scal^c,\cdot]) \\
			- & \mu^c[s,u] \sum_{u^\prime \in  \Ucal_D^{c\star}(\mu)}  \rho^c_{uu^\prime}(F^c(\mu),\mu^c[\Scal^c,\cdot]).
		\end{split}
	\end{equation*}
	Notice that the first term is null because, by hypothesis, $\mu^c[s,u^\prime] = 0$ for all $s\in \Scal^c$ and all $u^\prime \notin  \Ucal_D^{c\star}(\mu)$; the second is null due to \eqref{eq:proof_aux_fr_null}; and the third and forth are null due to statement~(i).
\end{proof}

Since $\mu$ is a MSNE, it follows from the definition of MSNE in Definition~\ref{def:MSNE} that $\mu$ satisfies the conditions of Lemma~\ref{lem:pc_prop}, therefore, by  Lemma~\ref{lem:pc_prop}(ii), $f^{c,r}_{s,u}(\mu) = 0$ for all $c\in [C]$, all $s\in \Scal^c$, and all $u \in \Ucal^c_D$. Furthermore, from the definition of MSNE, $f^{c,d}_{s,u}(\mu) = 0$ for all $c\in [C]$, all $s\in \Scal^c$, and all $u \in \Ucal^c_D$. Finally, $\dot{\mu}^c[s,u] = f^{c,d}_{s,u}(\mu)  + f^{c,r}_{s,u}(\mu) = 0$ for all $c\in [C]$, all $s\in \Scal^c$, and all $u \in \Ucal^c_D$, therefore $\mu$ is a rest point of \eqref{eq:ODE_mu_ev}.

\vspace{-0.4cm}

\subsection{Proof of Theorem~\ref{th:rest_is_MSNE}}\label{sec:proof_rest_is_MSNE}
	By the definition of rest point of \eqref{eq:ODE_mu_ev}, $\dot{\mu}^c[s,u]$ is null for all $c\in [C]$, all $s\in \Scal^c$, and all $u\in \Ucal_D^c$. As a result, since for all $c\in [C]$ and all $u\in \Ucal^c_D$ by conservation of mass $\sum_{s\in \Scal^c}f^{c,d}_{s,u}(\mu) = 0$, it follows that $\sum_{s\in \Scal^c} \dot{\mu}^c[s,u] = \sum_{s\in \Scal^c}f^{r,c}_{s,u}(\mu) = 0$. It follows from Lemma~\ref{lem:pc_ns} that for all $c\in [C]$ and all $u\in \Ucal_D^c$ $\mu^c[\Scal^c,u]> 0 \implies F^c_u(\mu) \geq F^c_v(\mu)\; \forall v\in \Ucal^c_D$. Therefore, $\mu$ satisfies condition \eqref{eq:MSNE_u} in Definition~\ref{def:MSNE} of a MSNE. It also follows from Lemma~\ref{lem:pc_prop}~(ii) that $f^{c,r}_{s,u}(\mu) = 0$ for all $c\in [C]$, all $s \in \Scal^c$, and all $u\in \Ucal_D^c$. By the definition of rest point of \eqref{eq:ODE_mu_ev}, $\dot{\mu}^c[s,u] = f^{c,d}_{s,u}(\mu) + f^{c,r}_{s,u}(\mu)$ is null for all $c\in [C]$, all $s \in \Scal^c$, and all $u\in \Ucal^c_D$. One concludes that $f^{c,d}_{s,u}(\mu) = 0$ for all $c\in [C]$, all $s \in \Scal^c$, and all $u\in \Ucal^c_D$, so by Assumption~\ref{ass:markov_unique} $\mu$ satisfies condition \eqref{eq:MSNE_s} in Definition~\ref{def:MSNE} of a MSNE.%\notef{This only holds for an hybrid protocol if it does not have any imitative component since it requires Lemma~\ref{lem:pc_prop}.}

%\appendices
%\section{Proofs}
%\input{section/proof_schur.tex}
%\input{section/proof_inv_bound.tex}
%\input{section/proof_boundedness.tex}
%\input{section/proof_OCI_equiv.tex}
%\input{section/proof_feas_iff.tex}

%\section*{Acknowledgment}

\vspace{-0.4cm}
\section*{References}
\vspace{-0.4cm}
\bibliographystyle{IEEEtran}
\bibliography{../../../../../Publications/bibliography/parsed-minimal/bibliography.bib,../../../../../Papers/_bib/references-gt.bib,../../../../../Papers/_bib/references-c.bib}

%\theendnotes

%\vspace{-0.8cm}
%
%\begin{IEEEbiographynophoto}{Leonardo Pedroso} (Graduate Student Member, IEEE) is working towards the Ph.D.\ degree in the Department of Mechanical Engineering, Eindhoven University of Technology, The Netherlands. \end{IEEEbiographynophoto}
%
%\vspace{-1cm}
%
%\begin{IEEEbiographynophoto}{Andrea Agazzi} received a Ph.D. in Theoretical Physics from the University of Geneva in 2017. Since 2024, he has been an Associate Professor in the Mathematics Department at the University of Bern, Switzerland.
%\end{IEEEbiographynophoto}
%
%\vspace{-1cm}
%
%\begin{IEEEbiographynophoto}{W.P.M.H. Heemels} (Fellow, IEEE) received a Ph.D. (EE, control theory) degree (summa cum laude) from the Eindhoven University of Technology (TU/e) in 1999. Since 2006, he has been with the Department of Mechanical Engineering, TU/e, where he is currently a Full Professor and Vice-Dean. He is also an IFAC Fellow.
%\end{IEEEbiographynophoto}
%
%\vspace{-1cm}
%
%\begin{IEEEbiographynophoto}{Mauro Salazar} (Member, IEEE)  received the Ph.D. degree in Mechanical Engineering from ETH Z\"urich in 2019. He is currently an Associate Professor in the Control Systems Technology section at Eindhoven University of Technology.
%\end{IEEEbiographynophoto}

%\input{../bio/ieee_bio_pedroso.tex}
%\input{../bio/ieee_bio_agazzi.tex}
%\input{../bio/ieee_bio_heemels.tex}
%\input{../bio/ieee_bio_salazar.tex}

\end{document}